\documentclass[prd, aps, superscriptaddress, showpacs,nofootinbib]{revtex4-1}
\usepackage{amssymb,amsmath,amsthm,graphicx,amscd,mathtools}
\usepackage{upgreek,enumerate,color,verbatim,multirow,comment,ulem,slashed}
\usepackage[hidelinks]{hyperref}

\newcommand{\llangle}{\langle \hspace{-.8mm}\langle}  
\newcommand{\rrangle}{\rangle \hspace{-.8mm}\rangle}

\makeatletter                    
\@addtoreset{equation}{section}  
\makeatother                     

\begin{document}
\title{Fluctuation-Dissipation and Correlation-Propagation Relations from \\ the Nonequilibrium Dynamics of Detector-Quantum Field Systems}
\author{J.-T. Hsiang}
\email{cosmology@gmail.com}
\affiliation{Center for High Energy and High Field Physics, National Central University, Chungli 32001, Taiwan}
\author{B. L. Hu}
\email{blhu@umd.edu}
\affiliation{Maryland Center for Fundamental Physics and Joint Quantum Institute, University of Maryland, College Park, Maryland 20742-4111, USA}
\author{S. -Y. Lin}
\email{sylin@cc.ncue.edu.tw}
\affiliation{Department of Physics, National Changhua University of Education, Changhua 50007, Taiwan}

\begin{abstract}
We consider $N$ uniformly-accelerating Unruh-DeWitt detectors whose internal degrees of freedom are coupled to a massless scalar field in $(1+1)$D Minkowski space.  We use the influence functional formalism to derive  the Langevin equations  governing the  nonequilibrium dynamics of the internal degrees of freedom and show explicitly that  the system relaxes in time and equilibrates. We also show that once the equilibrium condition is established a set of fluctuation-dissipation relations (FDR) and correlation-propagation relations (CPR) emerges  for the detectors, extending earlier results of~\cite{RHA} which discovered these relations for the quantum field.  Although  similar in form to the FDRs commonly known from linear response theory, which assumes an equilibrium condition a priori, their physical connotations are  dissimilar from that of a nonequilibrium origin. We show explicitly that both sets of relations are needed to guarantee the balance of  energy flow in and out of the system in dynamical equilibrium with the field. 
These results are helpful to  investigations of quantum information and communications of detectors in space experiments and inquiries of theoretical issues in black holes and cosmology.  
\end{abstract}


\maketitle

\baselineskip=18pt
\allowdisplaybreaks

\newpage
\section{Introduction}



Fluctuation-dissipation relations (FDR) are fundamental relations established in statistical mechanics with wide-ranging implications in many areas of physics, theoretical and applied. The commonest form of FDR is often quoted in linear response theory (LRT) \cite{Kubo66,KuboBook} in the context of many-body or condensed matter systems \cite{KadBay,FetterWalecka}.  However FDRs can exist in a more general setting.  When viewed in the open systems or driven dissipative systems perspective the usual FDR in the LRT form emerges in a system interacting with its environment after it has relaxed to an equilibrium state at late times (e.g., \cite{QTD1}), or when kept in a nonequilibrium steady state (e.g. \cite{HHAoP}).  

As an example of FDR's  wide-ranging implications and applications, we mention the suggestion of Candelas and Sciama \cite{CanSci77,Sciama} in viewing a black hole  interacting with a quantum field as a dissipative system, and  viewing Hawking  \cite{Haw75}  radiation in light of a  FDR. (See also \cite{Mottola}.) Same interpretation of the physics applies to the Unruh effect \cite{Unr76} experienced by a uniformly accelerated detector.  Even though it was later shown \cite{HRS} that the fluctuations that these authors refer to are not the proper statistical mechanical noise of the environment (quantum field) and  the FDR they wrote down has a mismatch, thus not really addressing the backreaction of quantum fields on a black hole spacetime, this way of thinking  appealed to one of us enough to have motivated him to launch a systematic investigation into the statistical mechanical properties of quantum fields in the presence of accelerating detectors \cite{HM94,RHA,RHK}, moving particles and masses \cite{JH1,GH1,GHL},  and interpreting the backreaction of quantum field processes in black holes \cite{HRS,SRH,HuRou} and the early universe \cite{HuSin95,CamVer96,LomMaz97,CamHu98} in the light of FDRs.  

The detectors in the above studies are spatially localized objects with internal degrees of freedom coupled to quantum fields. 
 Similar  detector-field systems are also used in the newly emergent field of relativistic quantum information \cite{HLL} to examine theoretical issues like  environmental influences on quantum coherence~\cite{HLPRD} and entanglement~\cite{LinHu09,HHPRD,HHPLB} which are essential for projected applications in quantum communications and teleportation \cite{LCHqt}. They can represent not only atoms distributed in space, but also mirrors with some internal degrees of freedom describing their optical properties such as reflectivity (e.g., \cite{GBH13,SLH15,Lin18}), or even those degrees of freedom entering as basic constituents in a dielectric (e.g., \cite{BehHu11}). In this capacity the investigation of detector-field interactions in this paper carries as much importance in terms of experimental capabilities for relativistic quantum information as atom-field interactions have provided for the bounty achievements in atomic and optical physics. 

Since measurements carried out in quantum fields are usually more subtle and difficult \cite{Sorkin88}, our results for relations between atoms facilitate more accessible experimental possibilities. For this purpose we need to address both the quantum field-theoretical and the quantum thermodynamical aspects of the detector-field system. 

In this backdrop we now describe what have been achieved before on the important issues in the statistical mechanical properties of detector-field systems, and what are the new results in our present investigation. 


\subsection{Results from \cite{RHA}: Existence of FDR \& CPR}

A significant work before 1995 is that of Raine, Sciama and Grove (RSG) \cite{RSG}, the key points therein were summarized and other prior work described in  \cite{RHA}. There, Raval, Hu and Anglin used the worldline influence functional  to treat an arbitrary number of Unruh-DeWitt detectors, modeled as harmonic oscillators, in arbitrary  yet prescribed motion,  minimally coupled to a massless quantum scalar field in $1+1$ dimensions~\footnote{Note the definition of minimal vs derivative coupling varies in the literature: E.g. RHA \cite{RHA} defined $\dot{Q}\phi$ as minimal coupling, following QED, the same as is used here. But~\cite{SLH15} refers to $Q\phi$ as minimal coupling, while $\dot{Q}\phi$ or $Q\dot{\phi}$ as derivative coupling~\cite{Lin18}}. This coupling provides a positive definite Hamiltonian in 1+1 dimensions and is of the form of scalar electrodynamics describing the coupling of charged particles to an electromagnetic field in 2D.
The field and the system of detectors are 
not coupled to each other until the initial moment of interaction, and the initial state of the field is assumed to be the Minkowski vacuum.
They aimed at describing the averaged effects of the scalar field on the dynamics of the system of detectors. 

In this  (`stochastic field theory') approach  based on open quantum system concepts and techniques (via the influence action) \cite{CalHu88,CalHu08}, the detector (system) dynamics is obtained by coarse-graining (`integrating over') the quantum scalar field (environment) which interacts with all the detectors.  From the influence action they obtained a set of coupled Langevin equations for the detectors under the influence of the scalar field environment. These effective equations for the system nonetheless contain the full quantum dynamics of the field. 

We mention two examples studied by RHA which may be viewed as a prelude to our present investigation, (I) two  inertial oscillators kept  at some fixed spatial separation,   interacting indirectly via a common scalar field; (II) a uniformly accelerated detector (oscillator 1) and an inertial detector (oscillator 2).  RHA showed that there exist FDR 
relating the fluctuations of the stochastic forces on the detectors to the dissipative forces. They also discovered a related set of CPR
between the correlations of stochastic forces on different detectors and the retarded and advanced parts of the radiation mediated by them. 

In problem (I) of two inertial detectors RHA find that the field which mediates the two detectors modifies the impedance functions ($L_{ij}$  from (3.34) of \cite{RHA}) of both detectors.  
Drawing on the properties of inductance in an inductor $L$ and the analogy with antenna systems, they introduced {\it self and mutual impedance} for the description, where the self-impedance measures the backreaction of each detector on itself, and mutual impedance measures the backreaction of one on the other, namely,
the change in the response of one detector due to the fluctuations of the field in the vicinity of the other one. The dissipative properties of each detector are correspondingly altered due to the presence of the other detector.  The field fluctuations (noise) in this case are relatively trivial, and non-trivial effects can be ascribed mainly to the impedance functions.


Problem (II) deals with a uniformly accelerated detector and an inertial detector, the latter is called a probe because it does not backreact on the accelerated detector. 
In terms of applications, for a better understanding of the early universe cosmology and black hole information issues, this problem
is of interest because it is the setting for investigating cross-horizon correlations and entanglement \cite{LCH08,LinHu10} which may be useful for information retrieval (such as after perturbations left the de Sitter horizon and reenter the Friedmann-Robertson-Walker horizon).
Here the noise associated with the field fluctuations and the field correlations between the two trajectories play a dominant role. Since the probe cannot causally influence the accelerated detector, the dissipative features of this problem are relatively trivial. Most of the terms in the correlations of the stochastic force acting on the probe cancel each other, leaving behind a contribution that arises purely from field correlations across the horizon. RHA observed that this cancellation 
follows from the dissipative properties of the accelerated detector and its free uncoupled dynamics. It does not explicitly involve the fluctuations of the field. Instead, they point out that this cancellation comes about because 
of the existence of a \textit{correlation-propagation relation} (CPR). This set of more general relations between the correlations of various detectors and the radiation mediated by them is  derived  from the FDR for the accelerated detector.  Such a relation can be equivalently viewed as a construction of the free field two-point function for each point on either trajectory from the two-point function along the uniformly accelerated trajectory alone. The remaining terms, which contribute to the excitation of the probe, are shown to represent correlations of the free field across the future horizon of the accelerating detector. In this problem, the dissipative properties of either detector remain unchanged by the presence of the other. This happens because the probe cannot influence the accelerated detector. However, the stochastic  force acting on the probe plays a non-trivial role.

\subsection{This investigation: FDR and CPR between the detectors}

In this paper we consider the nonequilibrium dynamics of a system of $N$ uniformly accelerating Unruh-DeWitt detectors with no direct coupling  but  interacting indirectly  via a common quantum field.  The earlier work of RHA, which demonstrated the existence of a FDR and discovered the existence of a CPR for the quantum field, was based on an examination of the properties of the Green functions in the quantum field, specifically the {\it Hadamard's elementary function} which represents the noise in the environment,  and the {\it retarded Green function} which enters in the dissipation of the system.  For our present problem we need to 1) Consider whether and how the system equilibrates, as a precondition for the existence of these relations. This involves describing the relaxation dynamics of the system. We then ask: 
2) Does there exist a FDR between the detectors\footnote{Interestingly, in LRT, e.g~\cite{Kubo66} and its applications to problems in condensed matter physics~\cite{Lovesey} when the details of the bath are of no great concern. The FDR was primarily derived for the system.} as induced by the FDR in the quantum field? 3) Does there exist a CPR between the detectors as induced by the CPR in the quantum field? 4) If  so, does the CPR help in understanding some aspect of quantum information, specifically the quantum correlations and quantum entanglement in a detector-field system?

We focus on issues 1) -- 3) in this paper, in the set up when $N$ causally connected detectors   in 1+1 dimensions  all undergo uniform accelerating motion but with different accelerations. This is represented by $N$ Rindler trajectories all in the $R$ wedge with the same asymptotes. One curious feature of this set up is that in the weak coupling limit, the detectors at different positions will have different Unruh temperatures at late times while they are in equilibrium with the same quantum field. We shall study the properties of a tensor response function which embodies both the FDR and the CPR for both the field and the detectors. The investigation of issue 4) will be left to a future paper. 

The major findings in our investigation are as follows:

\noindent  1) We examine the {\it nonequilibrium dynamics} of the system during the relaxation process, and  show that the system's achievement  of  dynamical equilibration is a necessary condition for the existence of the system's FDR/CPR, (we can refer to their combination as a set of generalized FDR). \\
2) Under the state of dynamical equilibrium,  we identify the generalized FDR of the system.  Thus we have extended the previous results on the FDR/CPR of the environment to those of the detectors. It is interesting to note that the former is formulated for the initial configuration of the environment, while the latter is only realizable in the final equilibrium state of the detectors.\\
3) The generalized FDR guarantees the {\it energy balance} between the detectors, and the balance between the detector and the quantum field. This offers a vivid physical interpretation of the generalized FDR. \\
4) The generalized FDR, obtained here from the fully nonequilibrium dynamics of the system, although similar in form to the FDR obtained from {\it linear-response theory}, has totally different physical meanings and  implications in quantum thermodynamics. \\

This paper is organized as follows. In Sec.~\ref{S:formalism}, we describe our model of $N$ Unruh-DeWitt detectors in uniform acceleration,  the field configurations, and the influence functional formalism  used  for our investigation of the nonequilibrium dynamics of the detectors' internal degrees of freedom. In Sec~\ref{S:example}  we use a simple example of two inertial detectors to illustrate how the FDR/CPR of the {\it detectors} can be obtained and how they are related to the FDR/CPR of the {\it  scalar field}. In Sec.~\ref{S:relaxation}  we examine the relaxation and equilibration  of the internal degrees of freedom of the $N$ detectors, then we define  and discuss the meaning  of the noise and dissipation kernels, and establish the FDR/CPR for the detectors. Then in Sec.~\ref{S:energy balance}  we show how the energy flows between the constituents of the system and those between the system and the environment come to a balance, after the equilibrium condition is reached. This energy balance is a physical embodiment  of the generalized FDR and a mathematical validation of the necessity of self-consistency in the formalism. 
In Appendix~\ref{S:appe} we highlight  a possible ambiguity in the definitions of the Hadamard's elementary function and the retarded Green's function of the system in Rindler space and   point out a subtle nature of temperature in the detector's response and the FDR between the system and its environment in our set up. Finally in Appendix~\ref{S:eteirwe}, for the purpose of comparison, we present a result in~\cite{RHA} on a generalized FDR of the {\it scalar field} for general timelike trajectories $z(t)$ and $z'(t)$ without horizons.

\section{Open Quantum System/ Influence Functional Formalism}\label{S:formalism}

Here in (1+1)D Minkowski spacetime, we consider $N$ identical Unruh-DeWitt detectors, whose internal degrees of freedom $Q_{i}$ (with $i=1$, $\cdots$, $N$) are modeled by simple harmonic oscillators in the detectors' internal spaces. Suppose the detectors uniformly accelerate from spatial infinity at the asymptotic past of Minkowski spacetime towards an inertial observer resting at the origin of the Minkowski coordinate system $x^{\mu}=(t,z)$ of our choice. This accelerating motion is most easily described by the Rindler coordinates~\cite{rindler, kazuhiro} $(\eta,\xi)$, which parametrize the part of Minkowski space, $z\geq\lvert t\rvert$, by
\begin{align}
	t&=\frac{e^{\mathsf{a}\xi}}{\mathsf{a}}\,\sinh\mathsf{a}\eta\,,&z&=\frac{e^{\mathsf{a}\xi}}{\mathsf{a}}\,\cosh\mathsf{a}\eta\,,
\end{align}
The  line element of Rindler space $ds$ is given by 
\begin{equation}\label{E:dveyhds}
	ds^{2}=-e^{2\mathsf{a}\xi}\bigl(d\eta^{2}-d\xi^{2}\bigr)\,.
\end{equation}
with $-\infty<\eta$, $\xi<\infty$, and the constant parameter $\mathsf{a}>0$. In terms of these Rindler coordinates $(\eta,\xi)$, the $i$-th detectors follows a coordinate line of fixed $\xi=\xi_i$ with a proper acceleration 
\begin{equation}
  \alpha(\xi_i)=\mathsf{a}\,e^{-\mathsf{a}\xi_i}
\end{equation}
in such a way that at $\eta=0$, passing through the $z$ axis of the Minkowski coordinate, they are all in closest proximity to the inertial observer at $z=0$.

In the comoving frame of the $i^{\text{th}}$ detector~\cite{detector}, the dynamics of oscillator $Q_i$ of mass $m$ and natural frequency $\omega$ in its internal space is described by the action
\begin{equation}
	S_{Q}[Q_i]=\int\!d\tau_i\;\Bigl[\frac{m}{2}\,\dot{Q}_i^{2}-\frac{m\omega^{2}}{2}\,Q_i^{2}\Bigr]\,,
\end{equation}
where an overdot denotes differentiation with respect to the proper time $\tau_i$ of the oscillator. In the background spacetime  with metric $g_{\mu\nu}$ in \eqref{E:dveyhds},  with its inverse and determinant denoted by $g^{\mu\nu}$ and $g$ respectively,  a massless scalar field $\phi$ described by the action
\begin{equation}
	S_{\phi}[\phi]=-\frac{1}{2}\int\!d^{2}x\sqrt{-g}\;g^{\mu\nu}\nabla_{\mu}\phi\nabla_{\nu}\phi
\end{equation}
is initially in its vacuum state, according to the inertial observer.   This scalar field couples only to the internal degree of freedom of the detector. To avoid infrared divergences in a lower dimension space, we consider a coupling with $\phi$  in the form which involves 
$\dot{Q}_i$ instead of $Q_i$, i.e. in the interaction term,
\begin{align}
	S_{\textsc{int}}[Q_i,\phi]&=\int\!d^{2}x\sqrt{-g}\;j_i(x)\phi(x)\,,&j_i(x)&=\lambda_i\int\!d\tau_i\;\dot{Q}_i(\tau_i)\,\frac{\delta^{(2)}[x^{\mu}-z^{\mu}_i(\tau_i)]}{\sqrt{-g}}\,,  \label{j}
\end{align}
where $\lambda_i$ is the coupling constant and $z^{\mu}_i(\tau_i)$ 
is the prescribed trajectory depicting the uniformly accelerating motion of the $i$-th detector\footnote{A word on the nature of $Q$ and $z$ in the  model used here:  the external or mechanical degree of freedom (mdf)  $z$ is prescribed and non-dynamical. It does not couple to the field or the internal degree of freedom (idf) $Q$ of the detector by any direct or indirect means.  The internal degree  of freedom of the detector $Q$ will undergo nonequilibrium evolution but its dynamics will not affect the external motion $z$ of the detector.  In a model like what is used in \cite{GBH13,SLH15} where  $z$ is dynamical there will be interplay between the idf  and the mdf through the field.}. Thus $j(x)$ can be understood as a source current from the internal degree of freedom of the detector. 
Combining the above elements, the full action for $N$ uniformly accelerating detectors is given by
\begin{align}
	S 
	&= S_\phi [\phi] + \sum_{i=1}^{N} \Bigl( S_Q[Q_i] + S_{\textsc{int}}[Q_i,\phi]\Bigr).
\end{align}
Note that we have chosen identical mass $m$ and natural frequency $\omega$ for all the detectors while the detector-field coupling constant $\lambda_i$ may be different for each detector. The fixed locations of the detectors $z_{i}^{1}(\tau_{i})=\xi_i$ in the Rindler coordinates have been assigned accordingly to get rid of the resonance in (1+1)D models \cite{Lin18}, so the late-time state of the detectors will be independent of the initial state.

\subsection{influence functional, coarse-grained and stochastic effective action} 
We adopt the framework of  Feynman-Vernon influence functional~\cite{feyn}  or Schwinger-Keldysh~\cite{sch, kel, ctp, qbm} (or  `in-in' or `closed time path') formalisms. The  closed-time-path coarse-grained effective action (equivalent to the influence action) $S_{\textsc{cg}}$ \cite{feyn,HPZ,HHPRD,QTD1} is  given by~
\begin{align}
	S_{\textsc{cg}}&=\sum_{i}\int\!d\tau_{i}\;\Bigl[m\,\dot{\Delta}_{i}^{(Q)}(\tau_{i})\dot{\Sigma}_{i}^{(Q)}(\tau_{i})-m\omega^{2}\,\Delta_{i}^{(Q)}(\tau_{i})\Sigma_{i}^{(Q)}(\tau_{i})\Bigr]\notag\\
	&\qquad\qquad\quad+\frac{1}{2}\int\!d^{2}x\sqrt{-g}\int\!d^{2}x'\sqrt{-g'}\;\Delta^{(j)}(x)\,G_{R}^{(\phi)}(x,x')\,\Sigma^{(j)}(x')\notag\\
	&\qquad\qquad\qquad\qquad\qquad\quad+\frac{i}{2}\int\!d^{2}x\sqrt{-g}\int\!d^{2}x'\sqrt{-g'}\;\Delta^{(j)}(x)\,G_{H}^{(\phi)}(x,x')\,\Delta^{(j)}(x')\,.  \label{Scg}
\end{align}
Hereafter, we will denote $\Delta^{(x)}$, $\Sigma^{(x)}$ by $\Delta^{(x)}=x^{(+)}-x^{(-)}$ and $\Sigma^{(x)}=(x^{(+)}+x^{(-)})/2$ for any variable $x$, and the superscripts $(+), (-) $ indicate  respectively  the forward and backward time branches along which the associated variable is evaluated.

The retarded Green's function $G_{R}^{(\phi)}(x,x')$ and the Hadamard's elementary function $G_{H}^{(\phi)}(x,x')$  of the scalar field\footnote{Their physical interpretations in the influence functional formalism can be found in~\cite{HHPRD, QTD1}, for example.} are defined by 
\begin{align}
	G_{R}^{(\phi)}(x,x')&=i\,\theta(t-t')\,\operatorname{Tr}\Bigl(\bigl[\phi(x),\phi(x')\bigr]\,\rho_{\phi}\Bigr)\,,\\
	G_{H}^{(\phi)}(x,x')&=\frac{1}{2}\,\operatorname{Tr}\Bigl(\bigl\{\phi(x,\phi(x')\bigr\}\,\rho_{\phi}\Bigr)\,,\end{align}
where $\theta(t)$ is the unit-step function and $\rho_{\phi}$ in this case is the density matrix of the scalar field associated with its initial state. The retarded Green's function $G_{R}^{(\phi)}(x,x')$ of the field in fact is independent of the field state because it consists of a product of the unit-step function and the Pauli-Jordan function $G_{}^{(\phi)}(x,x')=i\,[\phi(x),\phi(x')]$, which then is the commutator of the field operator, a $c$-number. The latter, together with the Hadamard's elementary function, constitutes the Wightman function of the scalar field
\begin{equation}
	G_{>}^{(\phi)}(x,x')=i\,\langle\phi(x)\phi(x')\rangle=\frac{1}{2}\,G_{}^{(\phi)}(x,x')+i\,G_{H}^{(\phi)}(x,x')\,.
\end{equation}
The Wightman function of a massless scalar field in the Minkowski vacuum when expressed in terms of the Rindler coordinates is given by
\begin{align}
	G_{>}^{(\phi)}(x,x')&=i\int_{0}^{\infty}\!\frac{d\kappa}{2\pi}\;\frac{1}{2\kappa}\biggl\{\Bigl[\coth\frac{\pi\kappa}{\mathsf{a}}\,\cos\kappa(\Delta_{\eta}+\Delta_{\xi})-i\,\sin\kappa(\Delta_{\eta}+\Delta_{\xi})\Bigr]\biggr.\notag\\
	&\qquad\qquad\qquad\qquad+\biggl.\Bigl[\coth\frac{\pi\kappa}{\mathsf{a}}\,\cos\kappa(\Delta_{\eta}-\Delta_{\xi})-i\,\sin \kappa(\Delta_{\eta}-\Delta_{\xi})\Bigr]\biggr\}\,.\label{E:kghbssr}
\end{align}
It has the same form as the one in a thermal state with temperature
\begin{equation}
	T=\frac{\mathsf{a}}{2\pi}\,,
\end{equation}
This offers one way to see the Unruh effect. Here, $\Delta_{\eta}$ and $\Delta_{\xi}$ denote $\Delta_{\eta}=\eta-\eta'$ and $\Delta_{\xi}=\xi-\xi'$ respectively. Eq.~\eqref{E:kghbssr} allows us to immediately identify the corresponding Hadamard's elementary function and the retarded Green's function.  After preforming the temporal Fourier transformation over $\Delta_{\eta}$ according to
\begin{equation}
	\widetilde{f}(\kappa)=\int_{-\infty}^{\infty}\!d\Delta_{\eta}\;f(\Delta_{\eta})\,e^{+i\kappa\Delta_{\eta}}\,,
\end{equation}
we obtain a relation between the Fourier transforms of the Hadamard's elementary function and the retarded Green's function
\begin{equation}\label{E:dfbgrssdw}
	\widetilde{G}_{H}^{(\phi)}(\kappa)=\coth\frac{\pi\kappa}{\mathsf{a}}\,\operatorname{Im}\widetilde{G}_{R}^{(\phi)}(\kappa)\,.
\end{equation}
This is valid for {\it fixed} $\xi$ and $\xi'$ (for the general cases with $\tau$-dependent $\xi$ and $\xi'$, see Appendix~\ref{S:eteirwe}), and $\widetilde{G}_{H,R}^{(\phi)}(\kappa)$ in fact still depends on $\Delta_{\xi}$. The diagonal terms {($\xi=\xi'$)} of \eqref{E:dfbgrssdw} observe the conventional FDR, which relates the local effects of quantum fluctuations of the scalar field and through its back-action,  the  dissipation in the system with which the field interacts. The off-diagonal terms observe a less familiar relation, that  between the long-range nonlocal correlations of the field and the non-Markovian effects in the system mediated by the ambient field. This is the CPR termed  by Raval, Hu and Anglin \cite{RHA}.  Viewed together we may call   \eqref{E:dfbgrssdw}  a {\it generalized} FDR for the scalar field environment.

In the context of the oscillator-environment, the corresponding Hadamard's elementary function and the Pauli-Jordan function are often denoted by $\overline{\nu}$ and $\overline{\mu}$ respectively, and their sum gives the Wightman function, $Z=\overline{\mu}+i\,\overline{\nu}$, of the environment.

Substituting  the  current $j$  from  \eqref{j} into the  expression \eqref{Scg} for the coarse-grained effective action and making use of the Feynman-Vernon identity~\cite{feyn}, we obtain the stochastic effective action \cite{JH1} $S_{\text{eff}}$
\begin{align}
	S_{\text{eff}}&=\sum_{i}\int\!d\tau_{i}\;\biggl\{\Bigl[m\,\dot{\Delta}_{i}^{(Q)}(\tau_{i})\dot{\Sigma}_{i}^{(Q)}(\tau_{i})-m\omega^{2}\,\Delta_{i}^{(Q)}(\tau_{i})\Sigma_{i}^{(Q)}(\tau_{i})\Bigr]\biggr.\notag\\
	&\qquad\qquad\qquad+\biggl.\sum_{j}\lambda_{i}\lambda_{j}\int\!d\tau_{i}\,d\tau_{j}\;\dot{\Delta}_{i}^{(Q)}(x)\,G_{R}^{(\phi)}(z_{i},z_{j})\,\dot{\Sigma}^{(Q)}_{j}+\lambda_{i}\dot{\Delta}_{i}^{(Q)}(\tau_{i})\zeta_{i}(\tau_{i})\biggr\}\,,  \label{Sst}
\end{align}
where stochastic noise $\zeta_{i}$ obeys the Gaussian statistics
\begin{align}
	\llangle\zeta_{i}(\tau_{i})\rrangle&=0\,,
	&\llangle\zeta_{i}(\tau_{i})\zeta_{j}(\tau_{j})\rrangle&=G_{H}^{(\phi)}[z_{i}(\tau_{i}),z_{j}(\tau_{j})]\,.
\end{align}
where $\llangle\cdots\rrangle$ denotes the stochastic ensemble average performed over the  noise distribution, not the quantum expectation value~\cite{feyn}. One can say that the noise $\zeta_{i}$ encapsulates the quantum fluctuations of the scalar field at the location of the $i^{\text{th}}$ detector.

The variation of this effective action with respect to $\Delta^{(Q)}_{i}$ gives the Langevin equation of motion of $Q_{i}$ for each detector,
\begin{align}
	&&\frac{\delta S_{eff}}{\delta \Delta^{(Q)}_{i}}\bigg|_{\Delta^{(Q)}_{i}=0}&=0\,,\notag\\
	&\Rightarrow&m\ddot{Q}_{i}(\tau_{i})+m\omega^{2}\,Q_{i}(\tau_{i})+\lambda^{2}\sum_{j}\frac{d}{d\tau_{i}}\int\!d\tau_{j}\;G_{R}^{(\phi)}[z_{i}(\tau_{i}),z_{j}(\tau_{j})]\dot{Q}_{j}(\tau_{j})&=-\lambda\,\dot{\zeta}_{i}(\tau_{i})\,,\label{E:bfsrdf}
\end{align}
where we have let $\lambda_{i}=\lambda_{j}=\lambda$. Obviously, in \eqref{E:bfsrdf}, the noise $\zeta_{i}$ will add a stochastic component in the dynamics of the  internal degree of freedom of the detectors, and in turn  this random motion will induce nonlocal causal influences from one internal degrees of freedom to the others via the ambient quantum field (see, e.g., explanations in \cite{HHPRD}) in the form of the retarded Green's function of the field. This is in the nature of a {\it quantum radiation}~\cite{JH1}, different from the classical radiation emitted by a moving charge. The change in the state of the internal degree of freedom of one detector will be passed onto the other detectors by means of the retarded field, which in turn incurs subsequent disturbances to the other internal degrees of freedom. This causes further modifications in their retarded fields, which will propagate  to and affect other detectors. The process will continue on until all the internal motions settle down in an equilibrium state,  if it exists. In due course, this causal influence gradually builds up correlations between the internal degrees of freedom among spatially-separated detectors. At the same time, the back-action of this quantum radiation from each internal degree of freedom gives rise to a quantum reactive force~\cite{JohHu05} which damps out the motion of $Q_{i}$, with a  strength  proportional to the time variation of the internal degree of freedom. Therefore, the more violent the motion of the internal degree of freedom is, the stronger the quantum reactive force will be. This is conducive to enabling the system to evolve toward an equilibrium configuration. Inadvertently ignoring this contribution in the equation of motion  will bring about in general growing fluctuations in the physical observables of $Q$, giving unphysical predictions.

Before we proceed to solving the equation of motion, we observe that the retarded Green's function $G_{R}^{(\phi)}$ of the scalar field
\begin{equation}\label{E:gbffnsm}
	G_{R}^{(\phi)}(x,x')=\theta(\eta-\eta')\int_{0}^{\infty}\frac{d\kappa}{2\pi}\frac{1}{\kappa}\;\Bigl[\sin\kappa(\Delta^{(\eta)}+\Delta^{(\xi)})+\sin\kappa(\Delta^{(\eta)}-\Delta^{(\xi)})\Bigr]\,,
\end{equation}
is time-translationally invariant in Rindler time $\eta$, but it does not have such a nice property when expressed in terms of the detector's proper time. The fact that the $G_{R}^{(\phi)}$ is  {stationary allows} us to solve the equation of motion by means of the Laplace/Fourier transformation. Since the infinitesimal proper time interval $d\tau$ and the Rindler time interval $d\eta$ are related by $d\tau=e^{\mathsf{a}\xi}\,d\eta$, we can write the equation of motion \eqref{E:bfsrdf} as
\begin{equation}\label{E:rtuhbwest}
	Q''_{i}(\eta_{i})+\omega^{2}e^{2\mathsf{a}\xi_{i}}Q_{i}(\eta_{i})+\frac{\lambda^{2}e^{\mathsf{a}\xi_{i}}}{m}\sum_{j}\frac{d}{d\eta_{i}}\int_{0}^{\infty}\!d\eta_{j}\;G_{R}^{(\phi)}(\eta_{i}-\eta_{j},\xi_{i}-\xi_{j})\,Q'_{j}(\eta_{j})=-\frac{\lambda}{m}\,e^{\mathsf{a}\xi_{i}}\,\zeta'_{i}(\eta_{i})\,,
\end{equation}
where we have denoted the corresponding time derivative with respect to Rindler time $\eta$ by a prime, i.e., $dQ(\eta)/d\eta=Q'(\eta)$. If the coupling between the detector's internal degree of freedom and the field is switched on at $t=0$, the lower limits of the proper time integrals in \eqref{E:bfsrdf} or that of the Rindler time integral in \eqref{E:rtuhbwest} will be set to zero. At $\eta=0$, the internal degrees of freedom $Q_{i}$ of the detectors are not in equilibrium with the ambient scalar field, which is in a thermal state described by \eqref{E:kghbssr} at the moment when the coupling is switched on, so $Q_{i}$  follow nonequilibrium dynamics. It is then of interest to inquire whether this evolution will settle down to equilibrium and how it does so through a careful analysis of the relaxation process. Thus, for example,  we cannot a priori presume  energy conservation which is possible only under stationarity conditions such as warranted by  the existence of an equilibrium state.  

The physics will be more transparent when we isolate the local contribution from the nonlocal ones in \eqref{E:rtuhbwest}. Doing so, we arrive at
\begin{equation}\label{E:rtuhbwes}
	Q''_{i}(\eta_{i})+\frac{\lambda^{2}e^{\mathsf{a}\xi_{i}}}{2m}\,Q'_{i}(\eta_{i})+\omega^{2}e^{2\mathsf{a}\xi_{i}}Q_{i}(\eta_{i})+\frac{\lambda^{2}e^{\mathsf{a}\xi_{i}}}{2m}\sum_{j\neq i}\theta(\eta_{i}-\ell_{ij})\,Q'_{j}(t_{i}-\ell_{ij})=-\frac{\lambda}{m}\,e^{\mathsf{a}\xi_{i}}\,\zeta'_{i}(\eta_{i})\,,
\end{equation}
where $\ell_{ij}=\lvert\xi_{i}-\xi_{j}\rvert$. No frequency or mass renormalization is needed in this case. The nonlocal contributions, the fourth term, reduce to time-delay terms, reflecting the nature of causal influences. It represents history-dependent influences from the other detectors in the full course of their dynamical evolutions. The second term, resulting from the local contribution, describes the frictional force in the equation of motion. If we define $\gamma=\lambda^{2}/4m$ as the usual damping constant, then different detectors can induce different effective   frequencies and damping rates due to the factor $e^{2\mathsf{a}\xi_{i}}$ being location-dependent. This factor accounts for the redshift of energy transfer between detectors along different paths.

Next we use a simple example of two static detectors to see how the CPR between the detectors emerges and to illustrate the central idea of this paper.

\section{Two inertial detectors: Self and mutual impedances}\label{S:example}

Consider two static detectors at $z$ and $z'$ in $(1+1)$D Minkowski space, coupled to a massless scalar field $\phi$ initially in its vacuum state. The equations of motion for the internal degrees of freedom $Q_{1}$, $Q_{2}$ of the detectors of unit mass are
\begin{align}
	\ddot{Q}_{1}(t)+\omega_{1}^{2}\,Q_{1}(t)+\frac{e_{1}^{2}}{2}\,\dot{Q}_{1}(t)+\frac{e_{1}e_{2}}{2}\,\dot{Q}_{2}(t-\ell)&=-e_{1}\,\zeta_{1}(t)\,,\\
	\ddot{Q}_{2}(t)+\omega_{2}^{2}\,Q_{2}(t)+\frac{e_{2}^{2}}{2}\,\dot{Q}_{2}(t)+\frac{e_{1}e_{2}}{2}\,\dot{Q}_{1}(t-\ell)&=-e_{2}\,\zeta_{2}(t)\,,
\end{align}
assuming $i, j= 1, 2$ from \eqref{E:rtuhbwes}. Here $\omega_{i}$ is the natural frequency of the internal degree of freedom of the $i^{\mathrm{th}}$ detector and $e_{i}$ is its coupling constant with the scalar field,  assumed to be switched on at $t=0$.  The detectors are separated by $\ell=\lvert z-z'\rvert$. Thus $t-\ell$ gives the retarded time for any disturbance in the field generated by one detector to propagate to the other.

In the matrix form, the inhomogeneous solution of $\mathbf{Q}(t)=(Q_{1},Q_{2})^{T}$ is given by
\begin{equation}\label{E:ghfbeda}
	\widetilde{\mathbf{Q}}^{(\textsc{inh})}(\omega)=\widetilde{\mathbf{D}}_{2}(\omega)\cdot\widetilde{\mathbf{F}}(\omega)\,,
\end{equation}
where $\widetilde{\mathbf{Q}}(\omega)$ is the Fourier transform of $\mathbf{Q}(t)$, and $\mathbf{D}_{2}(t)$ is a homogeneous solution, satisfying $\mathbf{D}_{2}(0)=0$,  whose Fourier transform is given by
\begin{equation}\label{E:gbrhbfg}
	\widetilde{\mathbf{D}}_{2}(\omega)=\begin{pmatrix}-\omega^{2}-i\,\dfrac{e_{1}^{2}}{2}\,\omega+\omega_{1}^{2}&-i\,\dfrac{e_{1}e_{2}}{2}\,\omega\,e^{+i\omega\ell}\\-i\,\dfrac{e_{1}e_{2}}{2}\,\omega\,e^{+i\omega\ell}&-\omega^{2}-i\,\dfrac{e_{2}^{2}}{2}\,\omega+\omega_{2}^{2}\end{pmatrix}^{-1}\,.
\end{equation}
The force term $F_{i}(t)=-e_{i}\dot{\zeta}_{i}(t)$. Eq.~\eqref{E:ghfbeda} implies that the inhomogeneous solution $\mathbf{Q}^{(\textsc{inh})}(t)$ is
{\begin{equation}\label{E:nrnfsdw}
	\mathbf{Q}^{(\textsc{inh})}(t)=\int_{0}^{t}\!dt'\;\mathbf{D}_{2}(t-t')\cdot\mathbf{F}(t')\,.
\end{equation}}
We shall discuss  relaxation dynamics and the existence of dynamical equilibration in the sections that follow. Here we assume that the homogeneous solution of $\mathbf{Q}$ is negligible at late times. A more complete treatment of the nonequilibrium dynamics of $N$ inertial detectors and the thermodynamics of such a system is given in~\cite{QTD1}.

We introduce the Hadamard's elementary function of $Q$ by\footnote{Some freedom may exist in the choice of the two-point functions for the detector system. Please see the Appendix~\ref{S:appe} for more details.}
\begin{equation}\label{E:dvgedsfs}
	G_{H}^{(Q)}(t_{i},t_{j})\equiv \llangle Q_{i}(t_{i})Q_{j}(t_{j})\rrangle\,.
\end{equation}
Here $\llangle\cdots\rrangle$ denotes the stochastic ensemble average associated with the influence functional formalism~\cite{feyn}, but it can be shown with the help of the reduced density matrix of $Q$ that this expression is equivalent to half of the quantum expectation value of the anti-commutator of $Q$ when the internal degree of freedom takes the quantum operator form~\cite{HHAoP}. Therefore, in principle, this Hadamard's elementary function is defined in the same way as that of the scalar field.  We can show that at late times, the Hadamard's elementary function for the internal degrees of freedom of the detectors is given by
\begin{align}
	\bigl[\mathbf{G}_{H}^{(Q)}(t,t')\bigr]_{ij}&\simeq\int_{0}^{t}\!ds\int_{0}^{t'}\!ds'\;\Bigl[\mathbf{D}_{2}(t-s)\Bigr]_{ik}\Bigl[\mathbf{D}_{2}(t'-s')\Bigr]_{jl}\llangle F_{k}(s)F_{l}(s')\rrangle\notag\\
	&=e_{k}e_{l}\int_{0}^{t}\!ds\int_{0}^{t'}\!ds'\;\Bigl[\mathbf{D}_{2}(t-s)\Bigr]_{ik}\Bigl[\mathbf{D}_{2}(t'-s')\Bigr]_{jl}\,\frac{\partial^{2}}{\partial s\,\partial s'}\Bigl[\mathbf{G}_{H}^{(\phi)}(s-s')\Bigr]_{kl}\,.\label{E:bgiutwe}
\end{align}
In the limit $t$, $t'\to\infty$, Eq.~\eqref{E:bgiutwe} reduces to
\begin{align}\label{E:gkfjgkjrt}
	\lim_{t,t'\to\infty}\bigl[\mathbf{G}_{H}^{(Q)}(t,t')\bigr]_{ij}&=e_{k}e_{l}\int_{-\infty}^{\infty}\!\frac{d\omega}{2\pi}\;\omega^{2}\Bigl[\widetilde{\mathbf{D}}_{2}^{*}(\omega)\Bigr]_{ik}\Bigl[\widetilde{\mathbf{D}}_{2}(\omega)\Bigr]_{jl}\,\Bigl[\widetilde{\mathbf{G}}_{H}^{(\phi)}(\omega)\Bigr]_{kl}\,e^{-i\omega(t-t')}.
\end{align}
We observe that the Hadamard's elementary function $G_{H}^{(\phi)}(t-t')$ of the scalar field is given by
\begin{align}\label{E:urtberrr}
	G_{H}^{(\phi)}(x-x')=\frac{1}{2}\int_{0}^{\infty}\!\frac{d\omega}{2\pi}\,\frac{1}{2\omega}\Bigl(e^{+i\omega\ell}+e^{-i\omega\ell}\Bigr)\Bigl(e^{+i\omega\Delta_{t}}+e^{-i\omega\Delta_{t}}\Bigr)\,,
\end{align}
where $\Delta_{t}=t-t'$, so that its Fourier transform is given by
\begin{equation}\label{E:oelnrgj1}
	\widetilde{G}_{H}^{(\phi)}(\omega)=\operatorname{sgn}(\omega)\,\frac{1}{4\omega}\,\Bigl(e^{+i\omega\ell}+e^{-i\omega\ell}\Bigr)\,.
\end{equation}
In \eqref{E:oelnrgj1}, $\operatorname{sgn}(\omega)$ is the signum function with $\operatorname{sgn}(\omega)=\pm1$ when $\omega\gtrless0$ or zero otherwise.  On the other hand, the Fourier transform of the Pauli-Jordan function $G^{(\phi)}(t,t')$ is 
\begin{equation}
	\widetilde{G}^{(\phi)}(\omega)=\frac{i}{2\omega}\,\Bigl(e^{+i\omega\ell}+e^{-i\omega\ell}\Bigr)\,,
\end{equation}
from which we can construct the Fourier transform of the retarded Green's function by
\begin{equation}
	\widetilde{G}_{R}^{(\phi)}(\omega)=-i\int_{-\infty}^{\infty}\!\frac{d\omega'}{2\pi}\;\frac{\widetilde{G}^{(\phi)}(\omega')}{\omega'-\omega-i\,0^{+}}\,,
\end{equation}
such that its imaginary part is
\begin{equation}\label{E:oelnrgj2}
	\operatorname{Im}\widetilde{G}_{R}^{(\phi)}(\omega)=-\frac{i}{2}\,\widetilde{G}^{(\phi)}(\omega)=\frac{1}{4\omega}\,\Bigl(e^{+i\omega\ell}+e^{-i\omega\ell}\Bigr)\,.
\end{equation}
From \eqref{E:oelnrgj1} and \eqref{E:oelnrgj2} we can thus explicitly show the existence of  FDR/CPR, or a {\it generalized FDR for the scalar field} in its Minkowski vacuum
\begin{equation}\label{E:gbirtdw}
	\widetilde{G}_{H}^{(\phi)}(\omega)=\operatorname{sgn}(\omega)\,\operatorname{Im}\widetilde{G}_{R}^{(\phi)}(\omega)\,.
\end{equation}
This is  well known and is  a consequence of the time-ordering among various Green's functions. In addition, it holds for all fixed values of $z$ and $z'$ even when the corresponding spacetime points $(t,z)$ and $(t',z')$ are spacelike separated. Although the retarded Green's function of the massless scalar field is zero for a non-lightlike interval, it does not imply that the corresponding temporal Fourier transform is also zero.

Eq.~\eqref{E:oelnrgj1} implies that
\begin{equation}
	\widetilde{\mathbf{G}}_{H}^{(\phi)}(\omega)=\operatorname{sgn}(\omega)\,\begin{pmatrix}\dfrac{1}{2\omega}&\dfrac{1}{4\omega}\,\Bigl(e^{+i\omega\ell}+e^{-i\omega\ell}\Bigr)\\\dfrac{1}{4\omega}\,\Bigl(e^{+i\omega\ell}+e^{-i\omega\ell}\Bigr)&\dfrac{1}{2\omega}
	\end{pmatrix}\,.
\end{equation}
On the other hand, from \eqref{E:gbrhbfg}, we can explicitly show that
\begin{align}
	\Bigl[\widetilde{\mathbf{D}}_{2}(\omega)\Bigr]^{-1}-\Bigl[\widetilde{\mathbf{D}}_{2}^{*}(\omega)\Bigr]^{-1}=-i\,2\omega^{2}\begin{pmatrix}\dfrac{e_{1}^{2}}{2\omega}&\dfrac{e_{1}e_{2}}{4\omega}\,\Bigl(e^{+i\omega\ell}+e^{-i\omega\ell}\Bigr)\\\dfrac{e_{1}e_{2}}{4\omega}\,\Bigl(e^{+i\omega\ell}+e^{-i\omega\ell}\Bigr)&\dfrac{e_{2}^{2}}{2\omega}\end{pmatrix}\,.
\end{align}
Thus, \eqref{E:gkfjgkjrt} becomes
\begin{align}\label{E:jitrtr}
	\bigl[\mathbf{G}_{H}^{(Q)}(t-t')\bigr]_{ij}&=\frac{i}{2}\int_{-\infty}^{\infty}\!\frac{d\omega}{2\pi}\;\operatorname{sgn}(\omega)\,\Bigl[\widetilde{\mathbf{D}}_{2}^{*}(\omega)\Bigr]_{ik}\Bigl\{\Bigl[\widetilde{\mathbf{D}}_{2}(\omega)\Bigr]^{-1}-\Bigl[\widetilde{\mathbf{D}}_{2}^{*}(\omega)\Bigr]^{-1}\Bigr\}_{kl}\Bigl[\widetilde{\mathbf{D}}_{2}(\omega)\Bigr]_{lj}\,e^{-i\omega(t-t')}\notag\\
	&=\int_{-\infty}^{\infty}\!\frac{d\omega}{2\pi}\;\operatorname{sgn}(\omega)\,\operatorname{Im}\Bigl[\widetilde{\mathbf{D}}_{2}(\omega)\Bigr]_{ij}\,e^{-i\omega(t-t')}\,,
\end{align}		
where we have used the property that $\widetilde{\mathbf{D}}_{2}(\omega)$ is symmetric. Thus we arrive at 
\begin{equation}\label{E:gnrjfs}
	\widetilde{\mathbf{G}}_{H}^{(Q)}(\omega)=\operatorname{sgn}(\omega)\,\operatorname{Im}\widetilde{\mathbf{G}}_{R}^{(Q)}(\omega)\,,
\end{equation}
if we identify $\widetilde{\mathbf{G}}_{R}^{(Q)}(\omega)=\widetilde{\mathbf{D}}_{2}(\omega)$. The latter can be verified with the help of the density matrix formalism. {Eq.~\eqref{E:gnrjfs} is the generalized FDR for the internal degrees of freedom of the two static detectors when the environmental field is initially in its vacuum.}

Since \eqref{E:nrnfsdw} is a matrix equation and describes how $\mathbf{Q}$ responds to the noise force $\mathbf{F}$, we may make an analogy with impedance in elementary electromagnetism. That is, writing
\begin{align}
	\widetilde{Q}_{1}(\omega)&=L_{11}(\omega)\,\widetilde{F}_{1}(\omega)+L_{12}(\omega)\,\widetilde{F}_{2}(\omega)\,,&\widetilde{Q}_{2}(\omega)&=L_{21}(\omega)\,\widetilde{F}_{1}(\omega)+L_{22}(\omega)\,\widetilde{F}_{2}(\omega)\,,
\end{align}
we may interpret $L_{11}$ and $L_{22}$ as self-impedance and $L_{12}$, $L_{21}$ as mutual impedance~\cite{RHA} with
\begin{align}
	L_{ii}(\omega)&=\widetilde{\chi}_{i}(\omega)\Bigl[1+\frac{e_{1}^{2}e_{2}^{2}}{4}\,\omega^{2}\,e^{+i2\omega\ell}\,\widetilde{\chi}_{1}(\omega)\widetilde{\chi}_{2}(\omega)\Bigr]^{-1}\,,\\
	L_{12}(\omega)=L_{21}(\omega)&=+i\,\dfrac{e_{1}e_{2}}{2}\,\omega\,e^{+i\omega\ell}\,\widetilde{\chi}_{1}(\omega)\widetilde{\chi}_{2}(\omega)\Bigl[1+\frac{e_{1}^{2}e_{2}^{2}}{4}\,\omega^{2}\,e^{+i2\omega\ell}\,\widetilde{\chi}_{1}(\omega)\widetilde{\chi}_{2}(\omega)\Bigr]^{-1}\,,
\end{align}
with $\widetilde{\chi}_{i}(\omega)$ being
\begin{equation}
	\widetilde{\chi}_{i}(\omega)=\Bigl(-\omega^{2}-i\,\dfrac{e_{i}^{2}}{2}\,\omega+\omega_{i}^{2}\Bigr)^{-1}\,.
\end{equation}
Following the previous discussions, we obtain 
\begin{equation}
	\llangle Q_{2}(t),Q_{2}(t')\rrangle=\int_{-\infty}^{\infty}\!\frac{d\omega}{2\pi}\;\operatorname{sgn}(\omega)\,\operatorname{Im}L_{22}(\omega)\,e^{-i\omega(t-t')}\,,
\end{equation}
at late times after the motion of the internal degrees of freedom is fully relaxed. {This turns out to be Eq. (3.23) of \cite{RHA} after rewriting.}

Observe that  \eqref{E:gbirtdw} is the generalized FDR of the scalar field for fixed spatial coordinates $z$ and $z'$. In fact, in (1+1)D Minkowski space, we may take  advantage  of the null coordinates  to derive the generalized FDR of the scalar field for general timelike trajectory $z(t)$ and $z'(t)$ as long as they do not possess horizons~\cite{RHA}. For the sake of comparison, we adapt the relevant material in Appendix~\ref{S:eteirwe}.

Next, we will formally solve the equation of motion for $N$ uniformly accelerating detectors, and examine the energy flow in and out of each detector because the balance of the net energy flow will be an indicator that the system dynamics approaches  equilibrium.

\section{relaxation to equilibrium}\label{S:relaxation}

Following the procedures and arguments worked out for inertial detectors~\cite{QTD1}, we  expect the nonequilibrium dynamics of the system to show relaxation in the present setup, but there may be   complications  arising from the asymmetric configuration from location-dependent effective parameters expressed in Rindler time,  even though technically using Rindler time has a few advantages outlined earlier. On the other hand, from the aspect of physical measurements, it seems more sensible to use the proper time to express the observables and their time derivatives because that is the time an observer comoving with the detector keeps.

Thus, we will discuss energy balance in terms of physical quantities defined in the comoving frame of the detector, but express them by the Rindler coordinates when necessary. We write the Langevin equation in the comoving frame \eqref{E:bfsrdf} but in terms of the Rindler time as
\begin{align}\label{E:brvsuyr}
	m\,e^{-2\mathsf{a}\xi_{i}}\frac{d^{2}Q_{i}}{d\eta_{i}^{2}}(\eta_{i})+m\omega^{2}\,Q_{i}(\eta_{i})+\lambda^{2}e^{-\mathsf{a}\xi_{i}}\sum_{j}\frac{d}{d\eta_{i}}\int\!d\eta_{j}\;G_{R}^{(\phi)}[z_{i}(\eta_{i}),z_{j}(\eta_{j})]\frac{dQ_{j}}{d\eta_{j}}(\eta_{j})=-\lambda e^{-\mathsf{a}\xi_{i}}\frac{d\zeta_{i}}{d\eta_{i}}(\eta_{i})\,.
\end{align}
We can solve this equation, if we perform the Laplace transformation of $Q$ for a fixed $\xi$ according to
\begin{equation}\label{E:rbygfsds}
	\overline{Q}(s)=\int_{0}^{\infty}\!d\eta\;Q(\eta)\,e^{-s\eta}\,.
\end{equation}
This renders Eq.~\eqref{E:brvsuyr} to an algebraic equation. Note that the Laplace transform $\overline{Q}(s)$ of $Q$ is defined in terms of the Rindler time, and it has an implicit dependence on $\xi$, as will be seen clearly later. After performing the Laplace transformation, we cast \eqref{E:brvsuyr} into the form
\begin{align}
	\Bigl\{\Bigl(m\,e^{-2\mathsf{a}\xi_{i}}s^{2}+m\omega^{2}\Bigr)\delta_{ij}+\lambda^{2}e^{-\mathsf{a}\xi_{i}}s^{2}\Bigl[\overline{G}_{R}^{(\phi)}(s)\Bigr]{}_{ij}\Bigr\}\,\overline{Q}_{j}(s)=-\lambda\, e^{-\mathsf{a}\xi_{i}}\,s\,\overline{\zeta}_{i}(s)+\text{initial conditions}\,,
\end{align}
where on the righthand side, the terms which are associated with the initial conditions are not explicitly shown. The matrix notation $\bigl[\overline{G}_{R}^{(\phi)}(s)\bigr]{}_{ij}$ is a shorthand for $\overline{G}_{R}^{(\phi)}(s;\xi_{i},\xi_{j})$. The presence of the damping term, as seen in \eqref{E:rtuhbwes}, implies that at late times when  relaxation is {nearly} complete, the contributions from the initial conditions of $Q_{i}(\eta)$ will become exponentially small, thus negligible. The inhomogeneous solution, on the other hand, can be expressed by
\begin{equation}\label{E:ryyvdere}
	\overline{Q}_{i}^{(\textsc{inh})}(s)=D_{ij}^{(2)}(s)\,\Bigl\{-\lambda\, e^{-\mathsf{a}\xi_{j}}\Bigl[s\,\overline{\zeta}_{j}(s)-\zeta_{j}(0)\Bigr]\Bigr\}\,,
\end{equation}
with 
\begin{align}\label{E:bghfhvd}
	\mathbf{D}^{(2)}(s)&=\Bigl[m\,s^{2}\mathbf{A}^{2}+m\omega^{2}\,\mathbf{I}+\lambda^{2}s^{2}\mathbf{A}\cdot\overline{\mathbf{G}}_{R}^{(\phi)}(s)\Bigr]^{-1}\,,
\end{align}
and $A_{ij}=e^{-\mathsf{a}\xi_{i}}\,\delta_{ij}$, so that with Rindler time, the inhomogeneous solution of $Q_{i}$ takes the form
\begin{align}
	Q_{i}^{(\textsc{inh})}(\tau)=Q_{i}^{(\textsc{inh})}(\eta)=\int_{C}\!ds\;\overline{Q}_{i}^{(\textsc{inh})}(s)\,e^{s\eta}&=\int^{\eta}_{0}\!d\eta'\;D^{(2)}_{ij}(\eta-\eta')\Bigl[-\lambda\,e^{-\mathsf{a}\xi_{j}}\frac{d\zeta_{j}(\eta')}{d\eta'}\Bigr]\notag\\
	&=-\lambda\int^{\eta}_{0}\!d\eta'\;\Bigl[G_{R}^{(Q)}(\eta-\eta')\Bigr]_{ij}\,\frac{d\zeta_{j}(\eta')}{d\eta'}\,,\label{E:brividrwer}
\end{align}
where the closed contour $C$ encloses all the poles of $\mathbf{D}^{(2)}(s)$~\footnote{We note that $\mathbf{D}^{(2)}(z)$ may have poles in the wrong half plane of complex $z$ when the coupling constant is exceptionally large, when two neighboring detectors are exceedingly close or  when the resonance occurs~\cite{Lin18}. We shall avoid these extreme conditions  because finite-size effect or feedback-recoil effects to the external degrees of freedom need be considered.}. Here we see how $\xi_{i}$ implicitly enters in $Q_{i}$ due to the factor $A_{ij}$, and in addition, $\overline{\mathbf{G}}_{R}^{(\phi)}(s)$ still has dependence on $\xi_{i}-\xi_{j}$. In \eqref{E:brividrwer}, we introduced the retarded Green's function $G_{R}^{(Q)}(\eta-\eta')$ of the internal degree of freedom $Q$ by
\begin{equation}\label{E:dvgwasgg}
	\Bigl[G_{R}^{(Q)}(\eta_{i}-\eta_{j})\Bigr]_{ij}\equiv D^{(2)}_{ij}(\eta_{i}-\eta_{j})\,e^{-\mathsf{a}\xi_{j}}\,,
\end{equation}
which will be of great use in the context of showing the energy balance and proving the generalized FDR. Eq.~\eqref{E:brividrwer} will be the single most important expression in the following discussions. 

A few words about the matrix notations of various Green's functions: The $i$--$j$ component of $\mathbf{G}$ means that the Green's function $G$ is evaluated at a pair of spacetime points $(\eta_{i},\xi_{i})$ and $(\eta_{j},\xi_{j})$, that is, $G(\eta_{i},\xi_{i};\eta_{j},\xi_{j})$, while the $i$--$j$ component of $\overline{\mathbf{G}}$ for the Laplace transform or $\widetilde{\mathbf{G}}$ for the Fourier transform of $\mathbf{G}$ tells that they are evaluated for a pair of spatial coordinates $\xi_{i}$ and $\xi_{j}$. This convention will apply to the Green's functions of the field as well as those of the internal degrees of freedom of the detectors.

From the general solution of the Langevin equation \eqref{E:brvsuyr}, 
the two point function $G_{H}^{(Q)}(\eta_{i},\eta_{j})\equiv \llangle Q_{i}(\eta_{i})Q_{j}(\eta_{j})\rrangle$ in general is not invariant under Rindler-time translations 
due to the nonequilibrium evolution of the system. However,  in the course to relaxation, damping will suppress any dependence on initial conditions of $Q$ and quell the components which are not time-translationally invariant. Explicitly, at late times we can show from \eqref{E:brividrwer} that 
\begin{align}
	G_{H}^{(Q)}(\eta_{i},\eta_{j})&=\lambda^{2}\int^{\eta_{i}}_{0}\!d\eta'_{k}\int^{\eta_{j}}_{0}\!d\eta'_{l}\;\Bigl[G^{(Q)}_{R}(\eta_{i}-\eta'_{k})\Bigr]_{ik}\,\Bigl[G^{(Q)}_{R}(\eta_{j}-\eta'_{l})\Bigr]_{jl}\,\frac{\partial^{2}}{\partial\eta'_{k}\partial\eta'_{l}}\Bigl[G_{H}^{(\phi)}(\eta'_{k}-\eta'_{l})\Bigr]_{kl}\notag+\cdots\\
	&=\lambda^{2}\int\!\frac{d\kappa}{2\pi}\;\kappa^{2}\,\Bigl[\widetilde{G}^{(Q)*}_{R}(\kappa)\Bigr]_{ik}\,\Bigl[\widetilde{G}^{(\phi)}_{H}(\kappa)\Bigr]_{kl}\,\Bigl[\widetilde{G}^{(Q)T}_{R}(\kappa)\Bigr]_{lj}e^{-i\kappa(\eta_{i}-\eta_{j})}\label{E:hvjfhvyr}\\
	&\qquad\qquad\qquad\qquad\qquad\qquad\qquad+\text{exponentially small terms at late times}\,,\notag
\end{align}
where we have used the approximation suitable for large time $t\to\infty$,
\begin{align}\label{E:ybgrtrtasq}
	\int_{0}^{t}\!ds\;\mathbf{G}_{R}^{(Q)}(t-s)\,e^{-i\,\kappa s}&=e^{-i\,\kappa t}\biggl[\int_{\infty}^{\infty}\!d\Delta\;\mathbf{G}_{R}^{(Q)}(\Delta)\,e^{-i\,\kappa\Delta}-\int_{t}^{\infty}\!d\Delta\;\mathbf{G}_{R}^{(Q)}(\Delta)\,e^{-i\,\kappa\Delta}\biggr]\notag\\
	&=e^{-i\,\kappa t}\,\widetilde{\mathbf{G}}_{R}^{(Q)*}(\kappa)+\mathcal{O}(e^{-\alpha t})\,,
\end{align}
with $\alpha$ being some positive number to describe the generic decaying behavior of $\mathbf{G}_{R}^{(Q)}$ with time due to dissipative, reactive force in the Langevin equation, and that $\mathbf{G}_{R}^{(Q)}(\Delta)$ is a retarded function.

Eq.~\eqref{E:hvjfhvyr} immediately tells us that the integrand is the Fourier transformation of $G_{H}^{(Q)}(\eta_{i},\eta_{j})$, that is,
\begin{equation}
	\Bigl[\widetilde{G}_{H}^{(Q)}(\kappa)\Bigr]_{ij}=\lambda^{2}\kappa^{2}\,\Bigl[\widetilde{G}^{(Q)*}_{R}(\kappa)\Bigr]_{ik}\,\Bigl[\widetilde{G}^{(\phi)}_{H}(\kappa)\Bigr]_{kl}\,\Bigl[\widetilde{G}^{(Q)T}_{R}(\kappa)\Bigr]_{lj}\,,
\end{equation}
or,  in   matrix notation,
\begin{equation}
	\widetilde{\mathbf{G}}_{H}^{(Q)}(\kappa)=\lambda^{2}\kappa^{2}\,\widetilde{\mathbf{G}}^{(Q)*}_{R}(\kappa)\cdot\widetilde{\mathbf{G}}^{(\phi)}_{H}(\kappa)\cdot\widetilde{\mathbf{G}}^{(Q)}_{R}(\kappa)\,,
\end{equation}
since $\widetilde{\mathbf{G}}^{(Q)}_{R}(\kappa)$ is symmetric in its indices. Now we will use the matrix identity
\begin{align}\label{E:bgfhvrh}
	\mathbf{B}^{-1}\cdot\bigl(\mathbf{C}-\mathbf{B}\bigr)\cdot\mathbf{C}^{-1}&=\mathbf{B}^{-1}-\mathbf{C}^{-1}\,,&&\Rightarrow&\bigl(\mathbf{C}^{*}\bigr){}^{-1}\cdot\bigl(\operatorname{Im}\mathbf{C}\bigr)\cdot\mathbf{C}^{-1}&=-\operatorname{Im}\bigl(\mathbf{C}^{-1}\bigr)\,,
\end{align}
to write \eqref{E:hvjfhvyr} as
\begin{align}
	G_{H}^{(Q)}(\eta_{i}-\eta_{j})&=-\int\!\frac{d\kappa}{2\pi}\;\cosh\frac{\pi\kappa}{\mathsf{a}}\,\Bigl[\widetilde{G}^{(Q)*}_{R}(\kappa)\Bigr]_{ik}\,\Bigl\{\operatorname{Im}\Bigl[\widetilde{G}^{(Q)}_{R}(\kappa)\Bigr]^{-1}\Bigr\}_{kl}\,\Bigl[\widetilde{G}^{(Q)T}_{R}(\kappa)\Bigr]_{lj}\,e^{-i\kappa(\eta_{i}-\eta_{j})}\notag\\
	&=\int\!\frac{d\kappa}{2\pi}\;\cosh\frac{\pi\kappa}{\mathsf{a}}\,\Bigl[\operatorname{Im}\widetilde{G}^{(Q)}_{R}(\kappa)\Bigr]_{ij}\,e^{-i\kappa(\eta_{i}-\eta_{j})}\,.\label{E:orutfhvgw}
\end{align}
To arrive at \eqref{E:orutfhvgw}, we have used the generalized FDR of the scalar field \eqref{E:dfbgrssdw} in the matrix form
\begin{equation}\label{E:dbghdre}
	\widetilde{\mathbf{G}}^{(\phi)}_{H}(\kappa)=\cosh\frac{\pi\kappa}{\mathsf{a}}\,\operatorname{Im}\widetilde{\mathbf{G}}^{(\phi)}_{R}(\kappa)\,,
\end{equation}
and Eq.~\eqref{E:dvgwasgg}, which gives $\widetilde{\mathbf{G}}^{(Q)}_{R}(\kappa)=\widetilde{\mathbf{D}}^{(2)}(\kappa)\cdot\mathbf{A}$, or explicitly 
\begin{equation}
	\widetilde{\mathbf{G}}^{(Q)}_{R}(\kappa)=\Bigl[-m\,\kappa^{2}\mathbf{A}+m\omega^{2}\,\mathbf{A}^{-1}-\lambda^{2}\kappa^{2}\,\widetilde{\mathbf{G}}_{R}^{(\phi)}(\kappa)\Bigr]^{-1}\,,
\end{equation}
to link the noise kernel of the field $\widetilde{\mathbf{G}}^{(\phi)}_{H}(\kappa)$ with the dissipation kernel $\widetilde{\mathbf{G}}^{(Q)}_{R}(\kappa)$ of the internal degree of freedom,
\begin{equation}
	\cosh\frac{\pi\kappa}{\mathsf{a}}\,\operatorname{Im}\Bigl[\widetilde{\mathbf{G}}^{(Q)}_{R}(\kappa)\Bigr]^{-1}=-\lambda^{2}\kappa^{2}\,\widetilde{\mathbf{G}}^{(\phi)}_{H}(\kappa)\,.
\end{equation}
Eq.~\eqref{E:orutfhvgw} nicely shows that after the internal degree of freedom $Q$ of the detector reaches equilibrium,   a generalized FDR appears among the kernel functions of  $Q$,
\begin{equation}\label{E:dfbewe}
	\widetilde{\mathbf{G}}_{H}^{(Q)}(\kappa)=\cosh\frac{\pi\kappa}{\mathsf{a}}\,\operatorname{Im}\widetilde{\mathbf{G}}^{(Q)}_{R}(\kappa)\,,
\end{equation}
in addition to the generalized FDR established earlier for  the scalar field \eqref{E:dbghdre}.

The generalized FDR in \eqref{E:dfbewe} looks identical to the conventional form, obtained in the framework of  linear-response theory (LRT), for a quantum system in thermal equilibrium at the temperature $\mathsf{a}/2\pi$. However, their physical origins and contexts are quite distinct, so it is worth some elaborations of \eqref{E:dfbewe}. In the conventional form based on LRT, the FDR is established for a general nonrelativistic quantum system in contact with a thermal bath. The interaction between them is assumed vanishingly weak so that the system can be described by an equilibrium thermal state of the Gibbs form, and the backaction of the system on the bath can be ignored. The correlations established between different components of the system are due to their  direct interactions; the bath barely plays any role in this aspect. 

In contrast, Eq.~\eqref{E:dfbewe} is formulated in a nonequilibrium setting:  At a given initial time, the internal degrees of freedom of the detectors start interacting with the ambient quantum field in a given initial state,   in this case the Minkowski vacuum as reported by an inertial observer. Since in general the initial state is not an eigenstate of the Hamiltonian of the  detector-field combined, the reduced dynamics of the internal degree of freedom of the detectors will undergo nonequilibrium evolution,  unlike in the LRT where the system is assumed to remain close to  equilibrium throughout in a Gibbs' distribution, regulated by a bath with a set temperature or chemical potential. The setting of LRT requires that the coupling between the system and the bath is  ultra-weak.
Here, in  an open system setting, stochastic noise force from the quantum field  induces dissipation in the nonequilibirum dynamics of the reduced system.  For this Gaussian model under study the coupling strength between the system and the bath can be arbitrarily strong. As long as the aforementioned assumptions are not violated, in general we find that the final relaxed state of the reduced system is quite distinct from the Gibbs thermal state, even though the correlation function of the field in its initial state, formulated in terms of Rindler coordinates, looks thermal in (1+1)D Minkowski space.

From the previous construction, we see that the generalized FDR in this setup emerges only after the internal degrees of freedom reach a final equilibrium state,  which,  owing to the nonequilibrium evolution of the system, is totally different from its initial state. From these considerations, it is quite amazing that the FDRs \eqref{E:dfbewe} assume the same form as the conventional one based on LRT.  Here we would like to stress again that the proportionality factor in \eqref{E:dfbewe} is related to the {\it initial state of the field}, not the final state of the internal degree of freedom. In other words, the generalized FDR of the scalar field is formulated based on its initial configuration, while the generalized FDR of the detectors acted on by the field as their environment is only realizable in their {\it final equilibrium state}.

This same appearance of the generalized FDR may not be hard to understand. As has been pointed out earlier,  at late times the dynamics of the reduced system of the {detectors' internal degrees} of freedom is dictated by the environmental scalar field. This is  most clearly seen from \eqref{E:brividrwer}. 

Examining more closely the matrix relation \eqref{E:dfbewe},  whereas its diagonal elements,  obeying the conventional FDRs of the scalar field  \eqref{E:dfbgrssdw}, express the local balance between  dissipation and fluctuations of the internal degree of freedom of each detector, its off-diagonal elements obey  the CPR which  connect the  mutual non-Markovian influences among different detectors mediated by the ambient scalar field. 
In the {next section, we will show that these off-diagonal elements are essential in maintaining the energy balance between the reduced system and its environment after they {have reached} 
equilibrium.  Thus lies  the deeper  physical significance of the generalized FDR.

\section{energy balance}\label{S:energy balance}

We now consider the energy balance between the internal degrees of freedom of the system and the ambient scalar field they interact with. From the equation of motion \eqref{E:rtuhbwes}, it is easy to see that this interaction takes the form of  i) reactive force due to quantum radiation, ii) driving force arising from the noise in the quantum field, and iii) non-Markovian influence between detectors mediated by the field. They act as conduits to relay and distribute energy among detectors via the surrounding field. In the transient regime of the nonequilibrium evolution, net energy flows between each detector and its environment can be nonvanishing and varying in time. After the system has reached equilibrium and settled in a stationary state, the energy flow in and out of each detector should come to balance. We show explicitly that this is indeed the case.

Since in this (1+1)D model no renormalization of the mass or the  frequency in the equation of motion is necessary, we do not need to isolate the damping term from the delay term as is done in the case of inertial detectors in (1+3)D Minkowski spacetime~\cite{QTD1,HHPRD}.

We first take a look at the power delivered by the noise or stochastic force  $-\lambda\,d\zeta_{i}/d\tau_{i}$, from the righthand side of equation of motion \eqref{E:bfsrdf}, to the internal degree of freedom $Q_{i}$ of the $i^{\text{th}}$ detector,
\begin{equation}\label{E:vbhrytgwe}
	\mathcal{P}_{\zeta}^{(i)}(\tau_{i})=-\lambda\, \llangle\,\frac{d\zeta_{i}}{d\tau_{i}}(\tau_{i})\frac{dQ_{i}}{d\tau_{i}}(\tau_{i})\,\rrangle\,.
\end{equation}
(In this equation there is no summation over the index $i$.)  It is essentially the expectation value of a mechanical power in the comoving frame of the $i^{\text{th}}$ detector. At late times the velocity of the internal degree of freedom $dQ_{i}/d\tau_{i}$ is given by
\begin{align}
	\frac{dQ_{i}}{d\tau_{i}}&\simeq\frac{d}{d\tau_{i}}\int_{0}^{\eta_{i}(\tau_{i})}\!d\eta_{j}\;D^{(2)}_{ij}(\eta_{i},\eta_{j})\,\Bigl[-\lambda\,e^{-\mathsf{a}\xi_{j}}\,\frac{d\zeta_{j}}{d\eta_{j}}(\tau_{j})\Bigr]\notag\\
	&=-\lambda\,e^{-\mathsf{a}\xi_{i}}\,\frac{d}{d\eta_{i}}\int_{0}^{\eta_{i}}\!d\eta_{j}\;\Bigl[G_{R}^{(Q)}(\eta_{i}-\eta_{j})\Bigr]_{ij}\,\frac{d\zeta_{j}}{d\eta_{j}}(\eta_{j})\,,&\tau_{i}&\gg\gamma^{-1}\,,
\end{align}
using \eqref{E:brividrwer}. The power \eqref{E:vbhrytgwe} can then  be written as
\begin{align}\label{E:gbftefrse}
	\mathcal{P}_{\zeta}^{(i)}(\tau_{i})=\lambda^{2}\,e^{-2\mathsf{a}\xi_{i}}\int_{0}^{\eta_{i}}\!d\eta_{j}\;\Bigl[\frac{d}{d\eta_{i}}G_{R}^{(Q)}(\eta_{i}-\eta_{j})\Bigr]_{ij}\,\llangle\frac{d\zeta_{i}}{d\eta_{i}}(\eta_{i})\,\frac{d\zeta_{j}}{d\eta_{j}}(\eta_{j})\rrangle\,.
\end{align}
Since $\llangle\zeta_{i}(\eta_{i})\zeta_{j}(\eta_{j})\rrangle$ gives the Hadamard's elementary function of the scalar field in the Rindler frame
\begin{equation}
	\llangle\zeta_{i}(\eta_{i})\zeta_{j}(\eta_{j})\rrangle=G_{H}^{(\phi)}(\eta_{i}-\eta_{j},\xi_{i}-\xi_{j})\equiv\Bigl[G_{H}^{(\phi)}(\eta_{i}-\eta_{j})\Bigr]_{ij}\,,
\end{equation}
which depends on the difference of two Rindler times. We can make a change of variable $\eta =\eta_{i}-\eta_{j}$ in \eqref{E:gbftefrse} and arrive at
\begin{align}
	\mathcal{P}_{\zeta}^{(i)}(\tau_{i})=\lambda^{2}\,e^{-2\mathsf{a}\xi_{i}}\int_{0}^{\eta_{i}}\!d\eta\;\Bigl[\frac{d}{d\eta}G_{R}^{(Q)}(\eta)\Bigr]_{ij}\Bigl[-\frac{d^{2}}{d\eta^{2}}G_{H}^{(\phi)}(\eta)\Bigr]_{ij}\,.
\end{align}
Taking the limit $\tau_{i}$ or equivalently $\eta_{i}$ to $+\infty$ and noting that $G_{R}^{(Q)}(\eta)$ is the retarded Green's function of $Q$, we find that in this limit
\begin{align}
	\mathcal{P}_{\zeta}^{(i)}(\infty)&=\lambda^{2}\,e^{-2\mathsf{a}\xi_{i}}\int_{-\infty}^{\infty}\!d\eta\;\Bigl[\frac{d}{d\eta}G_{R}^{(Q)}(\eta)\Bigr]_{ij}\Bigl[-\frac{d^{2}}{d\eta^{2}}G_{H}^{(\phi)}(\eta)\Bigr]_{ij}\notag\\
	&=\lambda^{2}\,e^{-2\mathsf{a}\xi_{i}}\int_{-\infty}^{\infty}\!\frac{d\kappa}{2\pi}\;\kappa^{3}\Bigl[\operatorname{Im}\widetilde{\mathbf{G}}_{R}^{(Q)}(\kappa)\Bigr]_{ij}\Bigl[\widetilde{\mathbf{G}}_{H}^{(\phi)}(\kappa)\Bigr]_{ij}\,,\label{E:ngbhge1}
\end{align}
where  the real part of $\widetilde{\mathbf{G}}_{R}^{(Q)}(\kappa)$ is an even function of $\kappa$. The Fourier transform of $Q(\eta)$ can be related to the corresponding Laplace transform, defined in \eqref{E:rbygfsds}, by
\begin{equation}
	\widetilde{Q}(\kappa)=\overline{Q}(s=-i\kappa)\,,
\end{equation}
or a Wick rotation on the complex $s$ plane. Note that the expression in \eqref{E:ngbhge1} is time independent.

Likewise, the power delivered by the damping term and the non-Markovian effect is defined as
\begin{align}
	P_{\gamma}^{(i)}(\tau_{i})+P_{c}^{(i)}(\tau_{i})&=-\lambda^{2}\sum_{j}\frac{d}{d\tau_{i}}\int_{0}^{\infty}\!d\tau_{j}\;G_{R}^{(\phi)}[z_{i}(\tau_{i}),z_{j}(\tau_{j})]\,\llangle\,\frac{dQ_{j}}{d\tau_{j}}(\tau_{j})\frac{dQ_{i}}{d\tau_{i}}(\tau_{i})\,\rrangle\\
	&=-\lambda^{2}e^{-2\mathsf{a}\xi_{i}}\sum_{j}\int_{0}^{\eta_{i}}\!d\eta_{j}\;\Bigl[\frac{d}{d\eta_{i}}\,G_{R}^{(\phi)}(\eta_{i}-\eta_{j},\xi_{i}-\xi_{j})\Bigr]\Bigl[\frac{\partial^{2}}{\partial\eta_{i}\partial\eta_{j}}G_{H}^{(Q)}(\eta_{i},\eta_{j})\Bigr]\,.\notag
\end{align}
Earlier we have shown that at late times the Hadamard's elementary function $G_{H}^{(Q)}(\eta_{i},\eta_{j})$ of the internal degree of freedom $Q$
will become invariant under Rindler-time translations. 
Thus at late times this power will take the form
\begin{align}
	P_{\gamma}^{(i)}(\tau_{i})+P_{c}^{(i)}(\tau_{i})=-\lambda^{2}e^{-2\mathsf{a}\xi_{i}}\int_{0}^{\eta_{i}}\!d\Delta\;\Bigl[\frac{d}{d\Delta}\,G_{R}^{(\phi)}(\Delta)\Bigr]_{ij}\Bigl[-\frac{d^{2}}{d\Delta^{2}}\,G_{H}^{(Q)}(\Delta)\Bigr]_{ij}\,,
\end{align}
and in the limit $\tau_{i}\to\infty$ we obtain
\begin{align}
	P_{\gamma}^{(i)}(\infty)+P_{c}^{(i)}(\infty)&=-\lambda^{2}e^{-2\mathsf{a}\xi_{i}}\int_{-\infty}^{\infty}\!d\Delta\;\Bigl[\frac{d}{d\Delta}\,G_{R}^{(\phi)}(\Delta)\Bigr]_{ij}\Bigl[-\frac{d^{2}}{d\Delta^{2}}\,G_{H}^{(Q)}(\Delta)\Bigr]_{ij}\notag\\
	&=-\lambda^{2}e^{-2\mathsf{a}\xi_{i}}\int_{-\infty}^{\infty}\!\frac{d\kappa}{2\pi}\;\kappa^{3}\Bigl[\operatorname{Im}\widetilde{\mathbf{G}}_{R}^{(\phi)}(\kappa)\Bigr]_{ij}\Bigl[\widetilde{\mathbf{G}}_{H}^{(Q)}(\kappa)\Bigr]_{ij}\,.\label{E:ngbhge2}
\end{align}
Since Eq.~\eqref{E:dfbewe} says that the kernels $\widetilde{\mathbf{G}}^{(Q)}$ satisfy that same form of  generalized FDR as $\widetilde{\mathbf{G}}^{(\phi)}$ do, we immediately see that \eqref{E:ngbhge1} and \eqref{E:ngbhge2} add up to zero. That is, we have explicitly shown that 
\begin{equation}\label{E:rnlrnd}
	P_{\zeta}^{(i)}(\infty)+P_{\gamma}^{(i)}(\infty)+P_{c}^{(i)}(\infty)=0\,,
\end{equation}
for each detector after it has equilibrated and the energy balance is established. Physically this is a requisite condition for equilibrium. Mathematically, as is clearly demonstrated in our derivation, self-consistency in the formalism employed is an absolute necessity.

It is also worth mentioning that energy balance \eqref{E:rnlrnd} implies that
\begin{equation}
	\sum_{i}\Bigl[\frac{m}{2}\,\Bigl(\frac{dQ_{i}}{d\tau_{i}}\Bigr)^{2}+\frac{m\omega^{2}}{2}\,Q_{i}^{2}\Bigr]
\end{equation}
is a constant of motion. This is a stronger condition than what  the first integral of  the equation of motion \eqref{E:bfsrdf} will imply.


At this point, it is instructive to summarize the physics involved in the relaxation process of the detectors' internal degrees of freedom when the detectors undergo uniform  acceleration in an ambient massless scalar field. For the general initial configuration of the field, the internal degrees of freedom of the detectors will generically undergo nonequilibrium evolution with time. The noise in the quantum field generates a stochastic force, imparting a random component in the dynamics of the internal degrees of freedom of the detectors, which engenders quantum radiation. The back-action of this quantum radiation bring forth a reactive force which eases down their motion. These competitive processes are conduits of energy exchange between the detectors and their surrounding field,  enabling the equilibration. During the transient regime, the correlation between the internal degrees of freedom of the detectors in general is not stationary, but the nonstationary component is gradually suppressed by damping due to its  rapid  oscillating behavior. Thus, at late times, this correlation becomes stationary. This property is a key to the establishment of  a FDR for the internal degrees of freedom. Since we do not presume energy conservation, the emergence of the generalized FDR guarantees that 
the net energy flows between each detector and the ambient field, and those between the detectors, cease
at late times. In particular, without taking the CPR, that is, the relation governing the off-diagonal elements in the generalized FDR, into consideration, the energy balance cannot be achieved. The presence of energy balance signals the existence of an equilibrium configuration. This provides a physical understanding of the generalized FDR.

The formalism we employed produces self-consistent and exact results  for arbitrary coupling, including 
the strong coupling regime. In this case, as  stressed earlier, the final equilibrium state of  the internal degrees of freedom is not of the Gibbs form, so it is not straightforward to expect why the corresponding FDR still takes the form as \eqref{E:dfbewe} without going through some detailed analysis. Furthermore, at strong coupling, the interaction term in the total Lagrangian is not small and can be comparable with the detectors' components, so even if the equilibration is reached, we cannot exclude the possibility that a substantial portion of total energy might be hidden in the interaction term such that energy balance may take on a totally different form from \eqref{E:rnlrnd}, thus evading the neat expression for the generalized FDR. The analysis shown earlier precludes those possibilities.

\section{conclusion}

Fluctuation-dissipation relation (FDR) occupies a central place in statistical mechanics as it  succinctly captures the relation between the  dynamics of a (macroscopic) system and the (microscopic) fluctuations in its environment.  It offers an enlightened way to understand cosmological particle creation \cite{Par69} and Hawking-Unruh effects \cite{Haw75,Unr76} and, more importantly,   the  backreaction effects of  quantum field processes  on the dynamics of a  background spacetime as in a black hole \cite{CanSci77,Mottola} or the early universe \cite{HuSin95}.  This way of thinking drove earlier researchers (e.g., \cite{RHA,RHK}) to investigate into the relation between moving detectors and a quantum field, with the detector playing many roles, from an atom to a  mirror to even a black hole or the cosmos. Backreaction and fluctuation effects of quantum fields on curved spacetime are the central themes of semiclassical and stochastic gravity theories,  respectively \cite{HuVer08,HuVer20}, the two interim stages from quantum field theory in curved spacetime to quantum gravity. 

Another important motivation which  prompted us to continue this line of investigation begun in \cite{RHA} is the recent advent of relativistic quantum information, where detectors are used to probe into the quantum information contents and features of a quantum field. This is a fruitful line of inquiry as it is aided by more accessible atomic-optical experimental designs.

The model we studied is a system of $N$ uniformly accelerating Unruh-DeWitt detectors  whose internal degrees of freedom, modeled as harmonic oscillators, interact  with an   ambient quantum scalar field.  The oscillators have no direct coupling but can  indirectly interact  through the common field.   We follow the  nonequilibrium dynamics of this system and look for conditions for such relations to exist.  It is easy to see that in the transient regime when many time and  length scales compete no such relation exists.  But after the system has relaxed and equilibrated, 
both the FDR and the CPR relations for the system do exist.  We can refer to them collectively as the generalized FDR.  Indeed, the generalized FDRs exist for the {\it system} in addition to such relations already established earlier for the {\it environment} (quantum field) \cite{RHA}. Both sets of relations appear formally identical but their physical connotations are starkly different. These observations, together with an example we provide in the appendix, point out the subtle roles or interpretations of the temperatures  in the detectors' responses and in the generalized FDRs of the detectors and the field in the context of the Unruh effect.

Having shown that  a state of dynamical equilibrium results from the nonequilibrium dynamics of the internal degrees of freedom of the detectors we further investigated  the energy flow between the detectors and between each detector and the  field. The existence of {an energy-flow balance for a moving localized system} 
interacting with an  extended field in a curved background is  a nontrivial testimony of the power of  the generalized FDR. 
Since the late-time dynamics of the internal degrees of freedom of the detectors (system) 
is governed by the  quantum field (environment), 
it may not be hard to grasp the formal similarity between the generalized FDR of the system and that of the environment. 

Our next work is to extend this analysis to 1+3 dimensions \cite{HHLY}.  Thereafter we shall examine the dynamics for two detectors out of causal contact (such as the situation of Alice having fallen into a black hole and Bob staying outside of its event horizon \cite{LCH08}) and bring in quantum information considerations such as mutual information, purity, fidelity, decoherence and  entanglement \cite{LinHu10}. We shall use the  tools and results developed here and after for the analysis of information loss and retrieval in black holes and inflationary universes.\\

\noindent {\bf Acknowledgments} We thank Professors Chung-Hsien Chou and Kazuhiro Yamamoto  for helpful discussions. This work began when JTH visited BLH at the Maryland Center for Fundamental Physics of the University of Maryland.  BLH and JTH visited the National Center for Theoretical Sciences in Hsinchu, Taiwan during which interactions with SYL took place. The authors thank Prof. Chong-Sun Chu, the Director of NCTS,  for his kind invitation and hospitality. SYL is supported by the Ministry of Science and Technology of Taiwan under Grant No. MOST 106-2112-M-018-002-MY3 and in part by the National Center for Theoretical Sciences, Taiwan.

\appendix
\section{FDR-CPR and definitions of two-point functions}\label{S:appe}
We observe that there is more than one way to introduce the Hadamard's elementary function \eqref{E:dvgedsfs} for the internal degree of freedom of the detector. Thus here we explore whether the FDR and CPR may depend on different definitions of the two-point functions of $Q$.

Suppose, instead of \eqref{E:dvgwasgg}, we define $\widetilde{\mathbf{G}}^{(Q)}_{R}(\kappa)=\mathbf{A}^{-1}\cdot\widetilde{\mathbf{D}}^{(2)}(\kappa)$, that is, $\bigl[\widetilde{\mathbf{G}}^{(Q)}_{R}(\kappa)\bigr]_{ij}=e^{+\mathsf{a}\xi_{i}}\bigl[\widetilde{\mathbf{D}}^{(2)}(\kappa)\bigr]_{ij}$ such that 
\begin{equation}
	\widetilde{\mathbf{G}}^{(Q)}_{R}(\kappa)=\Bigl[-m\,\kappa^{2}\mathbf{A}^{3}+m\omega^{2}\,\mathbf{A}-\lambda^{2}\kappa^{2}\mathbf{A}\cdot\widetilde{\mathbf{G}}_{R}^{(\phi)}(\kappa)\cdot\mathbf{A}\Bigr]^{-1}\,,
\end{equation}
and define the Hadamard's elementary function $\mathbf{G}^{(Q)}_{H}(\eta_{i},\eta_{j})$ of $Q$ by
\begin{equation}\label{E:rubcqm}
	\Bigl[\mathbf{G}^{(Q)}_{H}(\eta_{i},\eta_{j})\Bigr]_{ij}=e^{\mathsf{a}\xi_{i}}\,\llangle Q_{i}(\eta_{i})Q_{j}(\eta_{j})\rrangle\,e^{\mathsf{a}\xi_{j}}\,.
\end{equation}
we can still obtain the same generalized FDR as \eqref{E:dfbewe} and maintain self-consistency in the balance of energy flow between each detector and its surrounding field.   Does this imply some ambiguities?   Here we take the stance of minimalism, that is, we choose to define the Hadamard's elementary function by \eqref{E:dvgedsfs} such that, in the language of quantum operators, it will be the quantum expectation values of the anti-commutator of the $\hat{Q}$ operators,  without additional factors as in \eqref{E:rubcqm}. Moreover, in the following, we will give an example to illustrate the advantage of the definition of $\mathbf{G}^{(Q)}_{H}$ in \eqref{E:dvgedsfs}.

Let us compute the velocity uncertainty $\llangle\dot{Q}^{2}(\tau)\rrangle$ of the oscillator in the internal space of the detector, which essentially is the proper-time derivative  of the Hadamard's elementary function of $Q$ in the coincident time limit. {It also signifies the mean kinetic energy of $Q$}. For simplicity we only consider the case of one uniformly accelerating detector, which spares us carrying the full machinery developed in earlier sections. Thus  we will not see non-Markovian effects due to field-induced interaction between detectors, but it does not affect our subsequent discussions.

From \eqref{E:rtuhbwes}, we see the equation of motion for a single detector coupled to the scalar field is given by
\begin{equation}
	Q''(\eta)+\omega^{2}\,e^{+2\mathsf{a}\xi}\,Q(\eta)+2\gamma\,e^{+\mathsf{a}\xi}\,Q'(\eta)=-\frac{\lambda}{m}\,e^{+\mathsf{a}\xi}\,\zeta'(\eta)\,.
\end{equation}
In principle, we may introduce the effective parameters such as
\begin{align}
	\Omega&=\omega\,e^{+\mathsf{a}\xi}\,,&\Gamma&=\gamma\,e^{+\mathsf{a}\xi}\,,&\Lambda=\lambda\,e^{+\mathsf{a}\xi}\,.
\end{align}
Again a prime denotes the derivative with respect to the Rindler time while an overdot represents the derivative with respect to the proper time of the detector.  In so doing, we obtain an equation of motion that looks like that of a typical Brownian oscillator, except that the parameters in the equation are position-dependent,
\begin{equation}
	Q''(\eta)+\Omega^{2}(\xi)\,Q(\eta)+2\Gamma(\xi)\,Q'(\eta)=-\frac{\Lambda(\xi)}{m}\,\zeta'(\eta)\,.
\end{equation}
We leave the physical interpretations of these position-dependent parameters aside for a moment, except for a comment that these effective parameter can be very large by themselves due to the exponential factor $e^{+\mathsf{a}\xi}$. This is nothing but the red/blue-shift factor, but discretion is advised when carrying out any perturbative arguments.

We solve this equation of motion for a fixed $\xi$. The standard procedures lead to
\begin{equation}
	Q(\eta)=\cdots-\frac{\Lambda}{m}\int_{0}^{\eta}\!d\eta'\;d_{2}(\eta-\eta')\,\zeta'(\eta')\,,
\end{equation}
with the Fourier transform of $d_{2}(\eta)$ given by
\begin{equation}
	\widetilde{d}_{2}(\kappa)=\Bigl[-\kappa^{2}+\Omega^{2}-i\,2\Gamma\,\kappa\Bigr]^{-1}\,,
\end{equation}
where $\cdots$ represents terms that depend on the initial conditions and are exponentially small at late times.

We would like to examine the asymptotic behavior of the coordinatevelocity uncertainty $\llangle Q'^{2}(\eta)\rrangle$, which will take the form
\begin{align}
	\llangle Q'^{2}(\eta)\rrangle&=\frac{\Lambda^{2}}{m^{2}}\int_{0}^{\eta}\!d\eta'\int_{0}^{\eta}\!d\eta''\;d'_{2}(\eta-\eta')d'_{2}(\eta-\eta'')\,\llangle\zeta'(\eta')\zeta'(\eta'')\rrangle\notag\\
	&=\frac{\Lambda^{2}}{m^{2}}\int_{0}^{\eta}\!d\eta'\int_{0}^{\eta}\!d\eta''\;d'_{2}(\eta-\eta')d'_{2}(\eta-\eta'')\,\frac{\partial^{2}}{\partial\eta\,\partial\eta'}G_{H}^{(\phi)}(\eta'-\eta'')\,,
\end{align}
where the Hadamard's elementary function of the massless scalar field is given by
\begin{equation}
	G_{H}^{(\phi)}(\eta'-\eta'')=\frac{1}{2}\int_{-\infty}^{\infty}\!d\kappa\;J_{\kappa}(\mathbf{x},\mathbf{x})\,\coth\frac{\pi\kappa}{\mathsf{a}}\,e^{-\kappa(\eta-\eta')}=\int_{-\infty}^{\infty}\!\frac{d\kappa}{2\pi}\;\widetilde{G}_{H}^{(\phi)}(\kappa)\,e^{-\kappa(\eta-\eta')}\,,
\end{equation}
with
\begin{equation}
	J_{\kappa}(\mathbf{x},\mathbf{x})=\biggl[2^{n-1}\pi^{\frac{n+1}{2}}\Bigl(\frac{e^{\mathsf{a}\xi}}{\mathsf{a}}\Bigr)^{n-2}\Gamma(\frac{n-1}{2})\biggr]^{-1}\frac{\sinh\frac{\pi\kappa}{\mathsf{a}}}{\mathsf{a}}\;\Gamma(\frac{n}{2}-1+i\,\frac{\kappa}{\mathsf{a}})\,\Gamma(\frac{n}{2}-1-i\,\frac{\kappa}{\mathsf{a}})\,,
\end{equation}
and in particular in $n=2$-dimensional spacetime,
\begin{equation}
	J_{\kappa}(\mathbf{x},\mathbf{x})=\frac{1}{2\pi\kappa}\,.
\end{equation}
Thus in the limit $\eta\to\infty$, the coordinate velocity uncertainty becomes
\begin{align}
	\llangle Q'^{2}(\infty)\rrangle&=\frac{\Lambda^{2}}{m^{2}}\int_{-\infty}^{\infty}\!\frac{d\kappa}{2\pi}\;\kappa^{4}\,\widetilde{d}_{2}^{*}(\kappa)\,\widetilde{d}_{2}(\kappa)\,\widetilde{G}_{H}^{(\phi)}(\kappa)=\frac{1}{m}\,e^{+\mathsf{a}\xi}\,\operatorname{Im}\int_{-\infty}^{\infty}\!\frac{d\kappa}{2\pi}\;\frac{\kappa^{2}}{-\kappa^{2}+\Omega^{2}-i\,2\Gamma\,\kappa}\,\coth\frac{\pi\kappa}{\mathsf{a}}\,.\label{E:rgbhveea}
\end{align}
We make a change of variable $\kappa=y\,e^{+\mathsf{a}\xi}$ and introduce a parameter $\alpha=\mathsf{a}\,e^{-\mathsf{a}\xi}$. Eq.~\eqref{E:rgbhveea} becomes
\begin{equation}
	\llangle Q'^{2}(\infty)\rrangle=\frac{e^{+3\mathsf{a}\xi}}{\pi^{2}m\mathsf{a}}\,\gamma\sum_{n=-\infty}^{\infty}\int_{-\infty}^{\infty}\!dy\;\frac{y^{4}}{[(\omega^{2}-y^{2})^{2}+4\gamma^{2}y^{2}][n^{2}+\frac{y^{2}}{\alpha^{2}}]}\,,\label{E:bghsvre}
\end{equation}
where we have used
\begin{equation}
	\coth\frac{\pi y}{\alpha}=\frac{y}{\pi\alpha}\sum_{n=-\infty}^{\infty}\frac{1}{n^{2}+\frac{y^{2}}{\alpha^{2}}}\,.
\end{equation}
We need to evaluate the integral in \eqref{E:bghsvre}
\begin{align}
	\int_{-\infty}^{\infty}\!dy\;\frac{y^{4}}{[(\omega^{2}-y^{2})^{2}+4\gamma^{2}y^{2}][n^{2}+\frac{y^{2}}{\alpha^{2}}]}&=\frac{\pi\alpha^{2}}{2\gamma}\biggl[\frac{\omega^{2}_{\textsc{r}}-\gamma^{2}}{(\lvert n\rvert\alpha+\gamma)^{2}+\omega^{2}_{\textsc{r}}}+\gamma\,\biggl(\frac{1}{\lvert n\rvert\alpha+\gamma+i\,\omega_{\textsc{r}}}+\frac{1}{\lvert n\rvert\alpha+\gamma-i\,\omega_{\textsc{r}}}\biggr)\biggr]\,,\label{E:gvdvreire}\
\end{align}
with the resonance frequency $\omega_{\textsc{r}}=\sqrt{\omega^{2}-\gamma^{2}}$, and then perform the subsequent summations
\begin{equation}
	\sum_{n=0}^{\infty}\frac{\pi\alpha^{2}}{2\gamma}\frac{\omega^{2}_{\textsc{r}}-\gamma^{2}}{(n\alpha+\gamma)^{2}+\omega^{2}_{\textsc{r}}}=i\,\pi\alpha\,\frac{\omega_{\textsc{r}}^{2}-\gamma^{2}}{4\omega_{\textsc{r}}\gamma}\Bigl[\psi(\frac{\gamma-i\,\omega_{\textsc{r}}}{\alpha})-\psi(\frac{\gamma+i\,\omega_{\textsc{r}}}{\alpha})\Bigr]\,,
\end{equation}	
where $\psi(z)$ is the Digamma function.

To evaluate the summation of the second term in the square brackets, we implement a cutoff regularization and obtain
\begin{align}
	\sum_{n=0}^{\infty}\frac{\pi\alpha}{2}\,\biggl[\;\frac{1}{n+\dfrac{\gamma+i\,\omega_{\textsc{r}}}{\alpha}}+\dfrac{1}{n+\dfrac{\gamma-i\,\omega_{\textsc{r}}}{\alpha}}\;\biggr]\,e^{-n\epsilon}&\simeq-\pi\alpha\Bigl(\gamma_{\varepsilon}+\ln\epsilon\Bigr)-\frac{\pi\alpha}{2}\,\biggl[\psi(\frac{\gamma-i\,\omega_{\textsc{r}}}{\alpha})+\psi(\frac{\gamma+i\,\omega_{\textsc{r}}}{\alpha})\biggr]\,,
\end{align}
with $\epsilon\to0^{+}$. Here we only have a logarithmic divergence. Since it is independent of $\alpha$, it is associated with the 
vacuum fluctuations of the field at zero temperature.

Thus the summation and the integral in \eqref{E:bghsvre} can be given by
\begin{equation}
	\sum_{n=0}^{\infty}\int_{-\infty}^{\infty}\!dy\;\frac{y^{4}}{[(\omega^{2}-y^{2})^{2}+4\gamma^{2}y^{2}][n^{2}+\frac{y^{2}}{\alpha^{2}}]}=-\frac{\pi\alpha^{3}}{2\omega_{\textsc{r}}\gamma}\,\operatorname{Im}\biggl\{\biggl(\frac{\gamma+i\,\omega_{\textsc{r}}}{\alpha}\biggr)^{2}\,\psi(\frac{\gamma+i\,\omega_{\textsc{r}}}{\alpha})\biggr\}+\cdots\,,
\end{equation}
and Eq.~\eqref{E:bghsvre} becomes
\begin{align}
	\llangle Q'^{2}(\infty)\rrangle&=\frac{e^{+3\mathsf{a}\xi}}{\pi^{2}m\mathsf{a}}\,\gamma\,\biggl[-\frac{\pi\alpha^{2}}{2\gamma}+2\sum_{n=0}^{\infty}\int_{-\infty}^{\infty}\!dy\;\frac{y^{4}}{[(\omega^{2}-y^{2})^{2}+4\gamma^{2}y^{2}][n^{2}+\frac{y^{2}}{\alpha^{2}}]}\biggr]\notag\\
	&=-\frac{\alpha\,e^{+2\mathsf{a}\xi}}{2\pi m}\,\biggl\{1+\frac{2\alpha}{\omega_{\textsc{r}}}\,\operatorname{Im}\biggl[\biggl(\frac{\gamma+i\,\omega_{\textsc{r}}}{\alpha}\biggr)^{2}\,\psi(\frac{\gamma+i\,\omega_{\textsc{r}}}{\alpha})\biggr]+\cdots\biggr\}\,.\label{E:dbgvyeta}
\end{align}
The $\dots$ represents the logarithmically divergent term, a contribution from the 
vacuum fluctuations of the field.

Now we will take some limits to get a better idea about what \eqref{E:dbgvyeta} delivers. In the limit $\alpha\to\infty$, we find
\begin{equation}\label{E:kghtjhd}
	\lim_{\alpha\to\infty}\llangle Q'^{2}(\infty)\rrangle=\frac{\alpha\,e^{+2\mathsf{a}\xi}}{2\pi m}\,\biggl[1+4\gamma_{\varepsilon}\,\frac{\gamma}{\alpha}+\mathcal{O}(\frac{1}{\alpha^{2}})\biggr]\simeq e^{+2\mathsf{a}\xi}\,\frac{1}{2m}\Bigl(\frac{\alpha}{\pi}\Bigr)+\cdots\,.
\end{equation}
Since $\alpha$ turns out to be the proper acceleration of the detector, we may identify the ratio $T_{\textsc{u}}=\alpha/\pi$ as the temperature perceived by the detector, that is, the local temperature the detector responds to. On the other hand, in the limit $\alpha\to0$, we have
\begin{equation}\label{E:dbgrytsd}
	\lim_{\alpha\to0}\llangle Q'^{2}(\infty)\rrangle\simeq e^{+2\mathsf{a}\xi}\,\frac{\gamma}{\pi m}\biggl[\ln\frac{\alpha^{2}}{\omega_{\textsc{r}}^{2}+\gamma^{2}}+\frac{\omega_{\textsc{r}}^{2}-\gamma^{2}}{\gamma\omega_{\textsc{r}}}\tan^{-1}\frac{\omega_{\textsc{r}}}{\gamma}\biggr]+\mathcal{O}(\alpha)\,.
\end{equation}
In addition, in the limit of weak coupling $\gamma\to0$, we have
\begin{align}
	\tan^{-1}\frac{\omega_{\textsc{r}}}{\gamma}&\to\frac{\pi}{2}\,,&\omega_{\textsc{r}}^{2}\pm\gamma^{2}&\to\omega^{2}
\end{align}
so that \eqref{E:dbgrytsd} reduces to
\begin{align}\label{E:urthrbfaa}
	\lim_{\gamma\to0}\lim_{\alpha\to0}\llangle Q'^{2}(\infty)\rrangle\simeq e^{+2\mathsf{a}\xi}\biggl[\frac{\omega}{2m}+\frac{\gamma}{\pi m}\ln\frac{\alpha^{2}}{\omega^{2}}\biggr]+\cdots\,.
\end{align}

The additional factor $e^{+2\mathsf{a}\xi}$ in \eqref{E:kghtjhd}--\eqref{E:urthrbfaa} results from the fact that here we compute the coordinate-velocity uncertainty defined with respect the the Rindler time $\eta$. When we express the result in terms of the proper time $\tau$, which is related to the Rindler time $\eta$ by $d\tau=e^{\mathsf{a}\xi}\,d\eta$, we arrive at
\begin{align}
	\llangle\dot{Q}^{2}(\infty)\rrangle=e^{-2\mathsf{a}\xi}\,\llangle Q'^{2}(\infty)\rrangle=\begin{cases}
		\dfrac{1}{2m}\Bigl(\dfrac{\alpha}{\pi}\Bigr)+\cdots\,,&\text{high temperature limit}\,,\vspace{6pt}\\
		\dfrac{\omega}{2m}+\dfrac{\gamma}{\pi m}\ln\dfrac{\alpha^{2}}{\omega^{2}}+\cdots\,,&\text{zero temperature limit}\,.
	\end{cases}
\end{align}
Thus we restore the familiar finite- and zero-temperature results, and this points out appropriate definitions of the kernel functions of the internal degree of freedom $Q$. It is also pointed out that the Unruh temperature $T_{\textsc{u}}$ measured by the detector is different from the fiducial temperature $T=\mathsf{a}/2\pi$ that appears in the correlation function of the scalar field \eqref{E:kghbssr} as well as in the generalized FDRs \eqref{E:dbghdre} and \eqref{E:dfbewe}. This difference is particularly interesting when there is more than one detector involved. This again points out the interesting nature of the generalized FDRs \eqref{E:dbghdre} and \eqref{E:dfbewe}. Finally, we observe that these temperatures satisfy the Ehrenfest-Tolman relation~\cite{Tol},  
\begin{equation}
	T_{\textsc{u}}\sqrt{-g_{\eta\eta}}=\text{constant}=T\,.
\end{equation}
That is, for a system in thermal equilibrium in curved space, different observers can see different coordinate temperatures. When we say a system is in thermal equilibrium, the temperature we refer to is the fiducial temperature.  Note this situation is fundamentally different from the physical temperature difference which allows for a system to settle in a nonequilibrium steady state, as described in~\cite{HHAoP}. 

This example illustrates the difference in temperature that appears in the response of each detector and in the formal expressions of the generalized FDR for the detectors, as well as for the  environmental field. In the Wightman function of the massless scalar field and, in particular, the generalized FDR of the detectors, we see the fiducial temperature appears in the expression. However, from the responses   of the detectors to the field,  the reading gives the Unruh temperature of each detector, which depends on the local proper acceleration. It implies that in the weak detector-field coupling limit, the final state of the internal degrees of freedom of  different detectors will be the canonical thermal state at different temperatures. In other words, different detectors at different acceleration $\alpha$ will be thermalized to different temperatures. These temperatures are related by the above relation.  In comparison, the static detectors at various locations in Minkowski space,  such as depicted in ~\cite{QTD1}, will give the same temperature reading corresponding to the vacuum state of the field. This raises an interesting question  in the uniform acceleration case about how we  experimentally measure or verify the existence of the generalized FDR.  Observe that the same parameter $\mathsf{a}$ appears in the generalized FDR  for the entire one-parameter family of accelerating detectors, one can invoke the concept of a  `bookkeeper'  used in \cite{TayWhe}. All of the observers comoving with the uniformly-accelerating detectors report back instantaneously their  findings about the state of the detectors to the bookkeeper.  He or she can use this record to determine the  parameter $\mathsf{a}$ which enters the generalized FDR. This is a reflection that the  notion of temperature varies with the observers, as long observed by Tolman, and that in a curved spacetime setting, one should add to it a red-shift factor $\sqrt{ g_{00}}$.  Here,  we encounter the equivalent situation (see Eq.(4.98) of \cite{BirDav}).

This issue is further compounded if we   allow  for strong interactions between the internal degrees of freedom and the field. Our treatment can  fully account  for this situation.  Earlier studies~\cite{LinHu06,LinHu07,QTD1} show that at strong coupling, after the internal degrees of freedom of the detector reach equilibrium, their end state   is in general not a Gibbs state. 
Furthermore, even if we can introduce an effective temperature for each uniformly accelerating detector , this temperature parameter will depend on  other parameters of the detectors, which, as pointed out earlier,  are position-dependent. This further obscures the physical meaning of the temperature for uniformly accelerated observers at strong coupling.

\section{generalized FDR of the field for general timelike trajectories}\label{S:eteirwe}

{In Sec.~\ref{S:example}, we identified the generalized FDR of the field for fixed but arbitrary spatial coordinates. In that case, since the field is in its vacuum state, the two-point functions of the field are stationary. This allows us to write the generalized FDR in a simple and familiar form in terms of the Fourier transforms of the two-point functions.}

{In fact, for the case of a massless scalar field in (1+1)D Minkowski space~\cite{RHA}, we can construct the FDR/CPR of the field for two moving detectors following {\it arbitrary timelike trajectories} $z_{i}(t)$ as long as they do not possess the horizons. However, since in this case, such constructed two-point functions of the field in general are not time-translationally invariant, the corresponding generalized FDR can only be expressed in the form of a convolution integral in time. Next we go over the derivation, based on \cite{RHA}, for comparison.}

Since the Pauli-Jordan function of the scalar field is given by
\begin{equation}
	G^{(\phi)}(t,z;t',z')=i\int_{0}^{\infty}\!\frac{d\omega}{2\pi}\;\frac{1}{2\omega}\Bigl[e^{+i\omega (u-u')}+e^{+i\omega (v-v')}-e^{-i\omega(u-u')}-e^{-i\omega(v-v')}\Bigr]\,,
\end{equation}
where the right- and the left-moving null coordinates $u$, $v$ are defined by $u=t-z$ and $v=t+z$ respectively,
we find that the relevant kernel function of the field in the Langevin equation, evaluated along the worldlines of the moving detectors is
\begin{align}
	\partial_{t}G^{(\phi)}[t_{i},z_{i}(t);t_{j},z_{j}(t_{j})]&=-\frac{1}{2}\int_{-\infty}^{\infty}\!\frac{d\omega}{2\pi}\;\Bigl[e^{+i\omega(v_{i}-v_{j})}\,\frac{dv_{i}}{dt_{i}}+e^{+i\omega(u_{i}-u_{j})}\,\frac{du_{i}}{dt_{i}}\Bigr]\notag\\
	&=-\frac{1}{2}\Bigl\{\delta[u_{i}(t_{i})-u_{j}(t_{j})]\,\frac{du_{i}}{dt}+\delta[v_{i}(t_{i})-v_{j}(t_{j})]\,\frac{dv_{i}}{dt}\Bigr\}\notag\\
	&=-\frac{1}{2}\Bigl\{\delta(t_{i}-t'_{j})+\delta(t_{i}-t''_{j})\Bigr\}\,,\label{E:dteutbd}
\end{align}
where we have introduced $t'_{j}=u^{-1}_{i}\circ u_{j}(t_{j}^{\vphantom{'}})$ and $t''_{j}=v^{-1}_{i}\circ v_{j}(t_{j}^{\vphantom{''}})$. On the other hand, if we evaluate the Hadamard's elementary function of the field along the worldlines of the detectors, we obtain from \eqref{E:urtberrr} that
\begin{align}\label{E:urtberrr2}
	G_{H}^{(\phi)}[t_{i},z_{i}(t_{i});t_{j},z_{j}(t_{j})]&=\int_{0}^{\infty}\!\frac{d\omega}{2\pi}\,\frac{1}{2\omega}\biggl\{\cos\omega\bigl[u_{i}(t_{i})-u_{j}(t_{j})\bigr]+\cos\omega\bigl[v_{i}(t_{i})-v_{j}(t_{j})\bigr]\biggr\}\notag\\
	&=\int_{0}^{\infty}\!\frac{d\omega}{2\pi}\,\frac{1}{2\omega}\biggl\{\cos\omega\bigl[u_{i}(t_{i}^{\vphantom{'}})-u_{i}(t'_{j})\bigr]+\cos\omega\bigl[v_{i}(t_{i}^{\vphantom{'}})-v_{i}(t''_{j})\bigr]\biggr\}\,.
\end{align}
Comparing this with \eqref{E:dteutbd}, we may introduce
\begin{align}
	\mu_{ij}^{(u)}(t,t')&=\int_{0}^{\infty}\!\frac{d\omega}{2\pi}\,\frac{1}{2\omega}\,\cos\omega\bigl[u_{i}(t)-u_{i}(t')\bigr]\,,&\gamma_{ij}^{(u)}(t,t')&=-\frac{1}{2}\,\delta[t-u_{i}^{-1}\circ u_{j}(t')]\,,\label{E:rgbhe1}\\
	\mu_{ij}^{(v)}(t,t')&=\int_{0}^{\infty}\!\frac{d\omega}{2\pi}\,\frac{1}{2\omega}\,\cos\omega\bigl[v_{i}(t)-v_{i}(t')\bigr]\,,&\gamma_{ij}^{(v)}(t,t')&=-\frac{1}{2}\,\delta[t-v_{i}^{-1}\circ v_{j}(t')]\,,\label{E:rgbhe2}
\end{align}
such that
\begin{align}
	\mu_{ij}^{(u)}(t_{i},t_{j})+\mu_{ij}^{(v)}(t_{i},t_{j})&=G_{H}^{(\phi)}[t_{i},z_{i}(t_{i});t_{j},z_{j}(t_{j})]\,,\\
	\gamma_{ij}^{(u)}(t_{i},t_{j})+\gamma_{ij}^{(v)}(t_{i},t_{j})&=\partial_{t}G^{(\phi)}[t_{i},z_{i}(t);t_{j},z_{j}(t_{j})]\,.
\end{align}
Essentially, $\mu_{ij}^{(u)}$, $\mu_{ij}^{(v)}$ are the left- and the right-moving components of the scalar-field Hadamard's elementary function along the detector trajectories. The definitions \eqref{E:rgbhe1}, \eqref{E:rgbhe2} imply that there exists a FDR/CPR of the field {in the time domain} among $G_{H}^{(\phi)}(t,t')$ and $\partial_{t}G^{(\phi)}(t,t')$ of the field along the worldlines of the two moving detectors,
\begin{equation}\label{E:kbgrwrada}
	\mu_{ij}^{(u,v)}(t,t')=\int_{-\infty}^{\infty}\!ds\;K^{(u,v)}(t,s)\,\gamma_{ij}^{(u,v)}(s,t')\,,
\end{equation}
for the right- and the left-moving components, with the transitive kernel function $K_{i}^{(u,v)}(t,t')$ defined by
\begin{align}
	K_{i}^{(u)}(t,t')&=-\int_{0}^{\infty}\!\frac{d\omega}{2\pi\omega}\;\cos\omega\bigl[u_{i}(t)-u_{i}(t')\bigr]\,,\\
	K_{i}^{(v)}(t,t')&=-\int_{0}^{\infty}\!\frac{d\omega}{2\pi\omega}\;\cos\omega\bigl[v_{i}(t)-v_{i}(t')\bigr]\,.
\end{align}
These relations \eqref{E:kbgrwrada}, although much more complex than \eqref{E:gbirtdw}, can be applied to the case that the detectors follow very general timelike trajectories without any event horizon. {In this sense, it is more general than \eqref{E:gbirtdw}, but it can not be extended to (1+3)D space.}


\end{document}